\documentclass[12pt]{article}

\usepackage{amsmath}
\usepackage{amsfonts}
\usepackage{lscape}
\usepackage{multicol}
\usepackage{a4p}
\usepackage{graphicx,graphics}
\usepackage{epsfig}
\def\thebibliography#1{\list    
 {[\arabic{enumi}]}{\settowidth\labelwidth{[#1]}\leftmargin\labelwidth
 \advance\leftmargin\labelsep
 \usecounter{enumi}}
 \def\newblock{\hskip .11em plus .33em minus -.07em}
 \sloppy
 \sfcode`\.=1000\relax}

\begin{document}  

\bibliographystyle{unsrt}

\begin{flushright}
February 5, 2001
\\
\end{flushright}


\vspace*{1cm}

\begin{center}
\Large
A Compilation of High Energy Atmospheric 
\\
Muon Data at Sea Level
\end{center}

\vspace*{3cm}

\begin{center}
Thomas Hebbeker
\\
Institut f\"ur Physik der Humboldt-Universit\"at zu Berlin,
\\
Invalidenstr. 110,  D-10115 Berlin, Germany
\end{center}

\begin{center} 
Charles Timmermans 
\\
Katholieke Universiteit Nijmegen,
\\
Toernooiveld 1,
N-6525 ED Nijmegen, Netherlands
\end{center}

\vspace*{3cm}


\bigskip


\bigskip

\bigskip

\begin{center}
\underline{ABSTRACT}
\end{center}

\bigskip


We collect and combine all published data on 
the vertical atmospheric muon flux and the muon charge ratio
for muon momenta above 10 GeV.
At sea level the world average of the  momentum spectra agrees with the 
flux calculated by E.V. Bugaev et al. within 15~\%. The observed shape of 
the differential flux versus momentum is 
slightly flatter than predicted in this 
calculation.
The experimental accuracy varies from 7\% at 10 GeV to 
17\% at 1 TeV. 
The ratio  of fluxes of positive to negative muons is found to be constant, 
at a value of $1.268$, with relative uncertainties increasing from  
approximately $1\%$ at low
momenta to about $6\%$ at 300 GeV.


\newpage

\underline{1. Introduction}

\bigskip


We collect measured atmospheric muon flux data and charge ratios 
as a function of momentum and compute world averages.
Only measurements  
at sea level 
or low altitudes 
and
for (near) vertical incidence are taken into account,
since several data sets are available for these experimental conditions. 
Here we consider only data with muon momenta above 10 GeV.
%
At lower momenta geomagnetic effects and solar 
influences play a significant role and make the interpretation of the
data more difficult. 
A recent compilation of charge ratio data 
at low muon momenta can be found in reference 
\cite{vulpescu}.


A precise knowledge of the muon spectrum and charge ratio 
allows to constrain the primary flux and the   
models of atmospheric showers so that also the atmospheric
neutrino fluxes can be calculated with a good precision.
This is a very important issue, since  
the Superkamiokande experiment\cite{sk} and others 
have seen indications for a disappearance of atmospheric muon 
neutrinos.
So far this interpretation is based on 
the angular distribution and on the ratio of muon neutrino to 
electron neutrino fluxes.
It is very important to compare also directly the 
measured and calculated absolute muon neutrino fluxes; 
until now this was prevented by the large model uncertainties.

\bigskip

\underline{2. Effects relevant for spectrum and charge ratio}

\bigskip

The following effects might influence the measurements of the
muon flux and the charge ratio.
It is possible that  
the published data need to be corrected accordingly 
in order to arrive at a
meaningful comparison between the various measurements. 

\begin{itemize}

\item \underline{Geomagnetic effects}
\\
For near vertical incidence the geomagnetic cutoff for primary
protons 
is below $10 \, \mathrm{GeV}$ for all latitudes 
at which the cosmic ray measurements were made\cite{honda}
(exceptions are discussed below).
%
Geomagnetic effects can therefore be 
neglected. 

\item \underline{Solar modulation}
\\
Using the parameterization given in reference \cite{nagashima}
we estimate that the primary proton flux 
at 50 GeV (100 GeV) decreases by $3\%$ (1.6 \%) 
at maximum solar activity compared
to the minimum.
The mean primary proton momentum resulting in 10 GeV 
muons at sea level exceeds $100 \, \mathrm{GeV}$.
Using the air shower program CORSIKA \cite{corsika}, we found that about 
$80\%$ of those protons have a momentum larger than $50 \, \mathrm{GeV}$. 
This results in an uncertainty of $\pm 1\%$ 
for the muon flux at a momentum of 10 GeV.
Similarly, one can estimate a flux uncertainty of $\pm 0.5\%$
at 20 GeV and less at higher momenta.
At 10 GeV the charge ratio is expected to change by about $\pm 0.2\%$. 
At higher momenta the effect is even smaller.
We do {\it not} correct the data for time dependent solar effects.

\item \underline{Altitude dependence}
\\
Not all experiments measure at sea level. In order to investigate
the dependence of flux and charge ratio
on the altitude we used the air shower simulation 
program CORSIKA
and also apply the empirical formula 
found by De Pascale et al.\cite{depascale1993}.

For muon momenta above 10 GeV and altitudes less than about 1000 m the 
vertical muon flux can be parameterized by
\begin{eqnarray}
   \frac{\Phi(h)}{\Phi(h=0)} =  e^{h/L} \pm 0.003
\end{eqnarray}
where 
$h$ = altitude,    
$ L = 4900 \, {\mathrm{m}} + 750 \, {\mathrm{m}} \,  
\frac{p}{\mathrm{GeV}}$ and    
$p$ = muon momentum.
   
The form of the parameterization is similar to the one used 
in\cite{depascale1993}.
The uncertainty of $\pm 0.003$ reflects the quality of the parameterization 
and the comparison to 
the measurements.
Example: For $h = 1000 \, \mathrm{m}$ and $p = 10 \, \mathrm{GeV}$
we obtain the flux 
$\Phi(h) = 1.08 \cdot  {\Phi(h=0)} $.
Note: Caprice data\cite{boezio2000} 
disagree with both \cite{depascale1993} 
and CORSIKA for higher momenta; here they have
not been taken into account.

The charge ratio is not affected, it changes 
by less than $0.005$ for $h < 1000 \, \mathrm{m}$ and 
$p > 10 \, \mathrm{GeV}$.

We do correct 
all published fluxes using formula (1).

\item \underline{Zenith angle dependence}
\\
The muon data are normally collected within a certain cone around the
vertical direction, including zenith angles up to $\theta^{max}$. 
With help of CORSIKA
we find that 
the zenith angle dependence can be 
parameterized in the form
\begin{eqnarray}
\frac{\mathrm{d} \Phi}{\mathrm{d} \cos \theta} 
\sim 1 + a(p) \cdot ( 1 - \cos \theta)
\end{eqnarray}
with a momentum dependent coefficient $a(p)$.
Accordingly we 
estimate the 
following flux reduction factors 
\begin{eqnarray}
g(\theta) = 
\frac{\mathrm{d} \Phi}{\mathrm{d} \cos \theta} (\theta) / 
\frac{\mathrm{d} \Phi}{\mathrm{d} \cos \theta} (0)
\end{eqnarray}

\begin{center}
\begin{tabular}{ccccc}
$p/\mathrm{GeV}$ & $a $ & $g(5^0)$ & $g(10^0)$ & $g(20^0)$ \\
\hline
 10   & -1.50  & 0.994   & 0.978 & 0.910  \\
 30   & -1.28  & 0.995   & 0.981 & 0.925  \\
 100  & -0.94  & 0.996   & 0.986 & 0.944  \\
 300  & -0.61  & 0.998   & 0.991 & 0.963  \\
 1000 & -0.22  & 0.999   & 0.997 & 0.987  \\
\end{tabular}
\end{center}

Note: the entries are differential values, they have not been 
integrated over $\theta$. 

Since not all experiments quote the range of accepted zenith angles, we 
{\it cannot} 
correct for this effect. We have to keep in mind that this 
might cause a bias,  
especially at low momenta.

\item \underline{Atmospheric pressure/temperature profile}
\\
Previous calculations\cite{barrett} and measurements\cite{macro}
indicate that the relative muon flux variation 
$\Delta \Phi$ 
at ground 
level is related to 
the temperature-distribution in the atmosphere via
\begin{eqnarray}
\frac{\Delta \Phi}{\Phi} 
= \alpha \cdot \frac{\Delta T_{eff}}{T_{eff}} 
\;\;\;\;  .
\label{eq1}
\end{eqnarray} 
$\Phi$ is the integral flux above a certain muon momentum threshold
$p_{th}$.
$T_{eff}$ is the absolute effective temperature of the 
higher atmosphere.
$\alpha$ is the temperature coefficient, which is a function of
zenith angle and muon energy. For zenith angles 
$\theta \approx 0$\cite{barrett,macro}:
\begin{eqnarray}
\alpha = 
\left[1 + 
\frac{70 \, \mathrm{GeV}}{p_{th}} \right]^{-1}
\label{eqalpha}
\end{eqnarray}
Simulations using CORSIKA arrive at similar conclusions.
Example: At a threshold of $70 \, \mathrm{GeV}$
the formula yields $\alpha = 0.5$. Since
the atmospheric temperature, with a typical value of 
$220 \, \mathrm{K}$, 
varies over the year by up to 
$\pm 5 \, \mathrm{K}$, this implies a flux change  
of $\pm 1 \%$.

For muon momenta above 10 GeV the pressure at ground level
is not expected to 
show
a significant correlation with the flux\cite{barrett}.

% ...
%

Unfortunately, most experiments do not report
the atmospheric temperature, nor do they correct for 
this effect. It is even not clear, how to define the 
reference value.
Therefore, we 
{\it cannot} 
correct for atmospheric effects.


\item \underline{Unfolding of the momentum spectrum}
\\
The measured muon spectrum agrees with the true 
spectrum only if the momentum resolution is small 
compared to the momenta being investigated. 
Otherwise, the steepness of the spectrum, which falls off approximately
according to
\begin{eqnarray}
\frac{\mathrm{d} \Phi}{\mathrm{d} p} 
\sim p^{-3}
\label{specapprox}
\end{eqnarray} 
leads to an asymmetric distortion, an enhancement of
the measured flux at high momenta.
Thus, the measured spectrum needs to be unfolded for
experimental resolution effects. In the most simple approach
- assuming the spectrum is roughly known - 
this can be achieved by a simple correction factor, which has
been calculated in \cite{hayman1962}.
The authors assume the spectrum (\ref{specapprox}) and 
a Gaussian error distribution 
in the variable $1/p$ with width
$\sigma_{1/p}$.
Often the experimental resolution is given in terms 
of the `Maximum Detectable Momentum' $p_{MDM}$, defined as the 
momentum value for which the integral over the Gauss distribution 
becomes $1/2$:
\begin{eqnarray}
 E(\frac{1/p_{MDM}}{\sqrt 2 \, \sigma_{1/p}})  = \frac{1}{2} 
\;\;\;\;\;\;
\mathrm{with} \;\;\;\;\;\;
E(x) \equiv  \frac{2}{\sqrt \pi} \, \int_0^x \, e^{-t^2} \; d \, t
\label{mdmdef}
\end{eqnarray} 
Thus, 
\begin{eqnarray}
 \frac{1}{p_{MDM}} =   0.6745 \; \sigma_{1/p}
\end{eqnarray} 
The ratio of the measured and true spectra 
is then given by
\begin{eqnarray}
 R\, (\frac{p_{MDM}}{p})  =  E(0.4769 \, \frac{p_{MDM}}{p}) + 
1.1829 \, \frac{p}{p_{MDM}} \, \exp 
\left(  - 0.2275 \, \frac{p_{MDM}^2}{p^2})  
\right)   
\label{rformula}
\end{eqnarray} 
The measured flux must be multiplied by $1/R$ to correct for 
the experimental resolution.
Figure~\ref{figunfold}
shows the dependence of R on $1/p$.
For $p < 0.3  \, p_{MDM} $ 
the correction amounts to less than $1\%$ and 
can be neglected. For higher momenta the correction rises
strongly and must be 
taken into account. 

\begin{figure}[htbp]
\begin{center}
\mbox{\epsfig{file=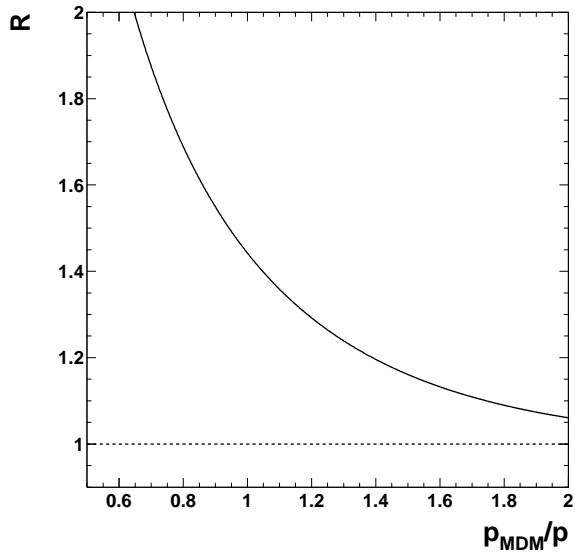,height=8cm,angle=0}}
\caption{Unfolding correction factor}
\label{figunfold}
\end{center}
\end{figure}

We have assumed that the experimenters have corrected their data 
for momentum resolution effects or that they can be neglected.
However, several  papers are not very clear on this point.
Therefore, some published spectra might be biased towards too high flux
values at large momenta. 


\end{itemize}

\bigskip

\underline{3. Experimental Data}

\bigskip

Only published results are taken into account. 
%
%
In appendix A we summarize the characteristics of  
all relevant experiments/publications, in chronological order.
The spectrum and charge ratio data used in this 
compilation are listed explicitly in appendices
B and C. 

\bigskip

\underline{4. Absolute muon flux}

\bigskip

There are two aspects to the measurement of the absolute muon flux, 
namely the shape of the spectrum as a function of energy and the absolute 
normalization. Some experiments only measure the relative muon flux as 
a function of momentum. Therefore we will analyze the data in two steps. 
First we check the spectral shape, leaving the normalization as a free 
parameter.
Secondly we determine
the absolute normalization of the spectrum.

\bigskip

\underline{4.1 The shape of the muon spectrum}

\bigskip

A whole range of experiments are performed to measure the muon flux, the
measurements used are listed in appendix B. 
We have corrected the datasets for altitude, 
which is a small correction in most cases. 
In order to be able to compare the
datasets, we fit each set to a reference shape, using the data 
with momenta above $10 \, \mathrm{GeV}$.
In this fit, and in the following, we 
assume that the measurement performed 
in each momentum bin is independent 
of the other momentum bins. The reference shape is taken from the 
theoretical calculation by Bugaev et al  \cite{bugaev}, leaving the 
normalization as a free parameter. In general this shape provides a good 
description of the datasets, as can be seen below.
The results of our fit are listed in table~\ref{shape}.
\begin{table}[h]
\begin{center}
\begin{tabular}{|l|c|c|c|}
\hline
Data set & $\rm \chi^2/NDF$ & Normalization & Normalization \\
& & from fit & from integration \\ \hline 
\hline
 Caro 1950 \cite{caro1950}         & 2.6/4  & $0.65 \pm 0.03$ & $0.74 \pm 0.09 $\\
 Owen 1955 \cite{owen1955}         & 0.5/2  & $0.819 \pm 0.013$ & 0.829 \\ 
 Pine 1959 \cite{pine1959}         & 4/11   & $0.76 \pm 0.03$   & 0.76 \\
 Pak 1961 \cite{pak1961}           & 4/6    & $0.75 \pm 0.03$   & 0.76 \\
 Holmes 1961 \cite{holmes1961}     & 43/12  & $0.807 \pm 0.016$ & 0.829 \\
 Hayman 1962 \cite{hayman1962s}    & 13/14  & $0.735 \pm 0.007$ & $0.746\pm 0.008$ \\
 Aurela 1967 \cite{aurela1967}     & 0.7/2  & $0.81 \pm 0.03$   & $0.79\pm 0.03$ \\
 Appleton 1971 \cite{appleton1971} & 38/23  & $0.370 \pm 0.003$ & $0.366\pm 0.003$ \\ 
 Allkofer 1971 \cite{allkofer1971} & 116/8  & $1.058 \pm 0.006$ & $1.01 \pm 0.01$ \\ 
 Bateman 1971 \cite{bateman1971}   & 8/8    & $0.871 \pm 0.008$ & $0.83 \pm 0.03$ \\
 Nandi 1972 \cite{nandi1972s}      & 60/14  & $0.998 \pm 0.008$ & $1.001 \pm 0.008$ \\
 Ayre 1975 \cite{ayre1975}         & 348/44 & $0.980 \pm 0.002$ & $0.95 \pm 0.02$ \\ 
 Green 1979 \cite{green1979}       & 2.3/4  & $0.98  \pm 0.02 $ & $0.98 \pm 0.02$ \\ 
 Rastin 1984  \cite{rastin1984a} (10-25 GeV)
                                   & 0.2/5  & $0.995 \pm 0.003$ & $0.977 \pm 0.002$ \\ 
 Rastin 1984  \cite{rastin1984a} ($>$25 GeV)
                                   & 24/29  & $0.960 \pm 0.005$ & $0.951 \pm 0.005$\\ 
 De Pascale 1993 \cite{depascale1993} 
                                   & 7/5    & $0.798 \pm 0.016$ & $0.80  \pm 0.03$ \\
 Tsuji  1998 \cite{tsuji1998}      & 16/13  & $0.961 \pm 0.014$ & $0.972 \pm 0.014$\\
 Kremer 1994 data \cite{kremer1999}& 10/6    & $0.822 \pm 0.009$ & $0.818 \pm 0.007$ \\ 
 Kremer 1997 data \cite{kremer1999}& 13/6    & $0.831 \pm 0.008$ & $0.821 \pm 0.007$\\ 
\hline
\end{tabular}
\end{center}
\caption{Normalization of datasets with respect to the Bugaev calculation }
\label{shape}
\end{table}
In this table, we separated the data from Rastin\cite{rastin1984a} 
into two sets, as different
normalizations are used in their paper. 
Next to fitting the normalization, we also
calculated the normalization by comparing the integrated flux above $\rm 10\,GeV/c$ 
(or the lower cutoff of the experiment whichever is higher) to an integrated flux
calculation using the reference shape. 
The data published in references  
\cite{owen1955}, \cite{holmes1961} and \cite{pak1961}  
do not allow for this normalization method. 
As these papers are normalized to Rossi \cite{rossi1948} we recalculate this
normalization point with the reference shape. 
The data of  \cite{bateman1971} are normalized 
to the differential flux at 10 GeV/c. 
The results of this calculation are shown in the last 
column of table~\ref{shape}. In general, both normalizations are in good agreement.\\
\begin{figure}[htbp]
\begin{center}
\mbox{\epsfig{file=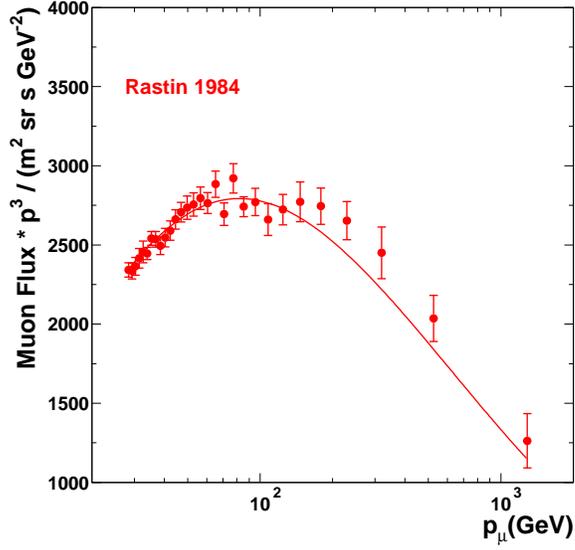,height=8cm,angle=0}}
\caption{Muon flux data by Rastin et al.\cite{rastin1984a} in comparison
to the reference spectrum from Bugaev et al.\cite{bugaev}, after
normalization}
\label{shrast}
\end{center}
\end{figure}
The high energy part of the Rastin data is shown in figure~\ref{shrast}. 
Here and in the following we present all spectra weighted with $p^3$, 
a common practice to compensate for the steep fall-off with momentum.
Figure~\ref{shrast} nicely
shows that the reference shape fits the data rather well, which justifies the use  
of the Bugaev curve as a reference. However,
the data has the tendency to be slightly higher than the normalized curve at the higher
momentum values. 
\\
The $\chi^2$ of the fit as listed in this table made us re-check 
five datasets; the first four are  
shown in figure~\ref{shholmes}. 
The data of Holmes\cite{holmes1961} 
clearly show
\begin{figure}[htbp]
\begin{center}
\mbox{\epsfig{file=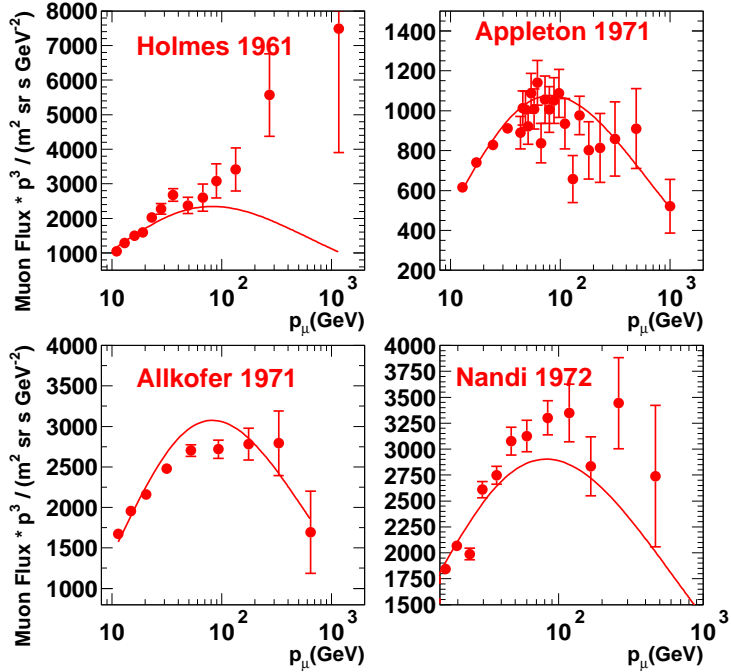,height=10cm,angle=0}}
\caption{Muon flux data by Holmes et al.\cite{holmes1961}, 
Appleton et al.\cite{appleton1971}, Allkofer et al.\cite{allkofer1971}
and Nandi et al.\cite{nandi1972} in comparison
to the reference spectrum from Bugaev et al.\cite{bugaev}, after
normalization}
\label{shholmes}
\end{center}
\end{figure}
that a simple re-normalization will not work. The
data points do not follow the reference shape, especially at higher momenta. In their 
paper Holmes et al. apply additional corrections to the highest two data points,
indicating that these are close to the MDM of the detector. 
Unfortunately, the value of this maximal
momentum is not mentioned. \\
The Appleton data\cite{appleton1971} are 
scattered a lot around the curve. With the value of $\chi^2/NDF$ being
only slightly less than 2, this plot suggests that 
the errors could be 
underestimated. \\
The Allkofer data\cite{allkofer1971} 
have a completely different shape. The data rise faster than the
reference shape and plateau at a lower value. 
This plateau also seems to be wider
than suggested by the reference distribution. \\
The Nandi data\cite{nandi1972} 
rise to a significantly higher value than predicted by the reference
shape. This and the low value of the third data point create the large   
$\chi^2$.
\begin{figure}[htbp]
\begin{center}
\mbox{\epsfig{file=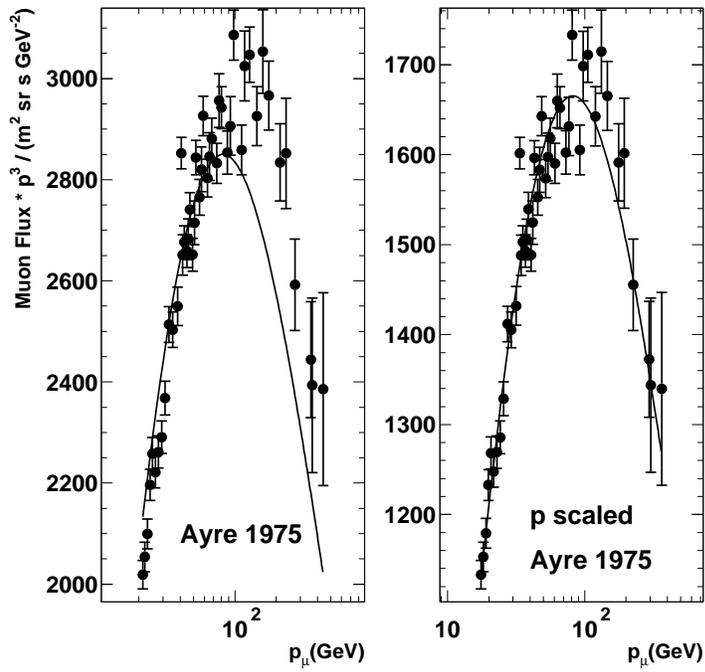,height=10cm,angle=0}}
\caption{Muon flux data by Ayre et al.\cite{ayre1975} in comparison
to the reference spectrum from Bugaev et al.\cite{bugaev}, after
normalization. Left: data as published. Right: spectrum after momentum
scaling.}
\label{shayre}
\end{center}
\end{figure}
\\
The Ayre data\cite{ayre1975}
(left side of fig. \ref{shayre}) 
start off below the curve and 
continue to rise longer than expected.
Therefore, the peak is at a higher value, but the drop-off rate seems to be similar 
as predicted on a log(p) scale. The curve suggests that the momentum could be 
over-estimated. A best fit of the momentum scale leads to a scaling 
of the momenta
by a factor of 0.825, see right side of fig. \ref{shayre}.
The $\chi^2/NDF$ improves from 348/44 to 136/44. It naturally 
changes the normalization. The muon spectrum 
closely follows  a $p^{-3}$-dependence, thus the 
normalization is changed to about 56 \% of the original. 
This is in fact what we observe.
Even using the modified momentum, for which we cannot find a justification, the Ayre 
data do not fit the curve very well.

\begin{figure}[htbp]
\begin{center}
\epsfig{file=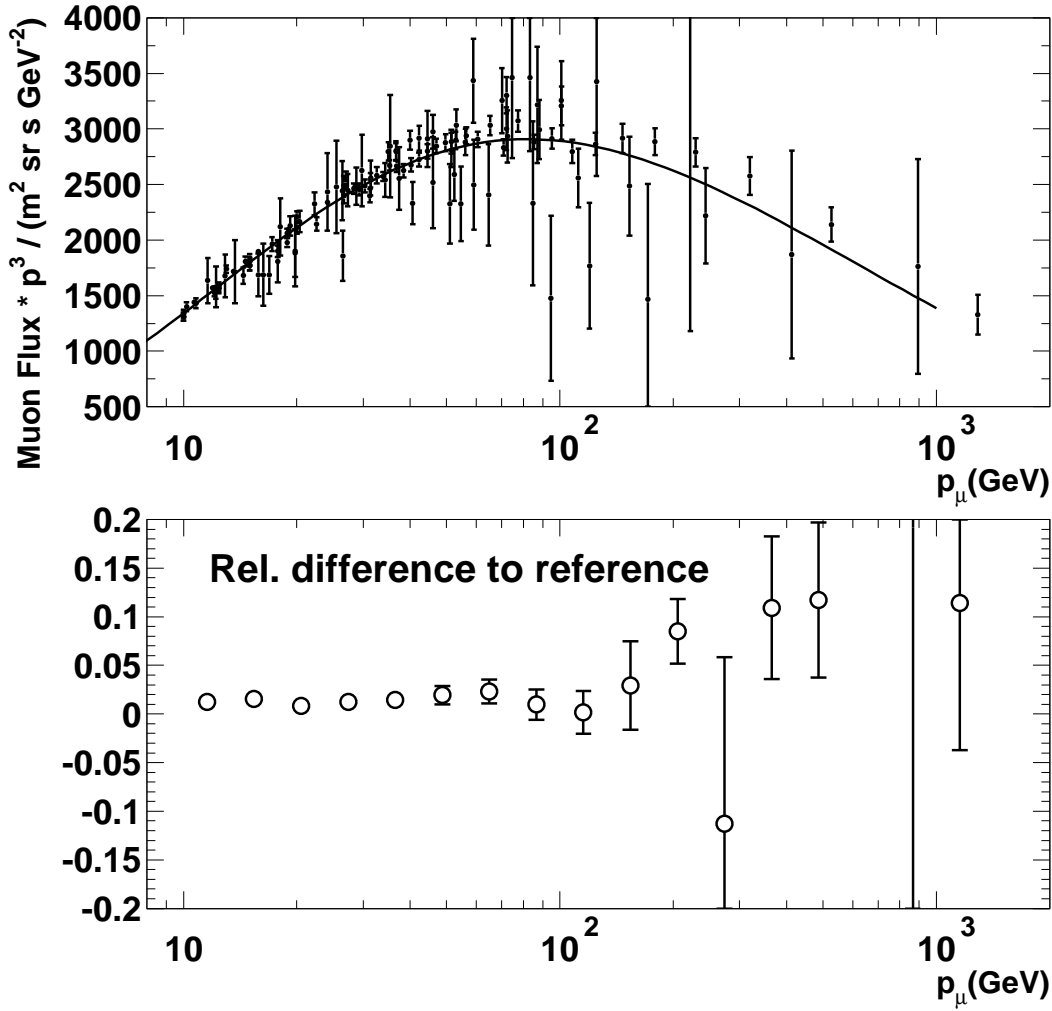,height=15cm,angle=0}
\caption{`Good' data in comparison to reference spectrum.}
\label{shgood}
\end{center}
\end{figure}

If we ignore these five datasets for the moment, we can compare the remaining data to
the reference curve. We do this by 
applying 
the normalization calculated as outlined before, and listed in 
table~\ref{shape} 
as `normalization from integration'. The result is shown in
figure~\ref{shgood}. 
\begin{figure}[htbp]
\begin{center}
\epsfig{file=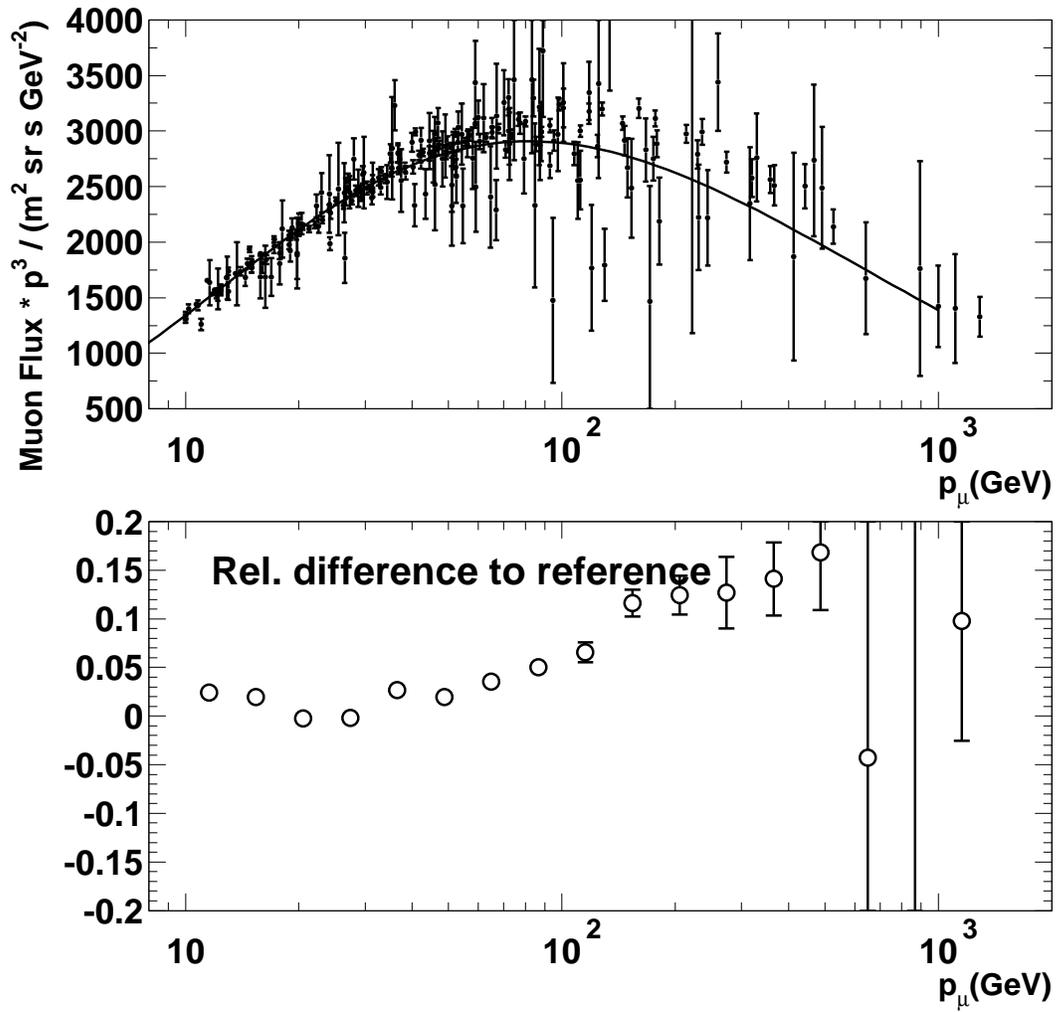,height=15cm,angle=0}
\caption{All data in comparison to reference spectrum.}
\label{shall}
\end{center}
\end{figure}
The top part of this plot shows a direct 
comparison of all the datasets to the theory, while the bottom part shows the relative difference between
these measurements and the description. 
The larger differences 
at higher momenta are mainly due to the data by 
Rastin et al. \cite{rastin1984a}. 
If we include the 
remaining five datasets 
the difference between the shape of the data
and the reference shape increases, especially at higher momenta. 
This can be
seen in figure~\ref{shall}. \\
The relative differences are plotted in 8 bins per decade, equidistant in 
log(p). 
This choice 
represents a good compromise taking into account
experimental uncertainties and the rate of change of the $p^3$-weighted 
spectrum with momentum.
The $\chi^2/NDF$ values and the average flux values 
in the units $\rm m^{-2}\, sr^{-1}\, s^{-1}\, GeV^{-1}$ 
are  shown for each bin in table \ref{shrat}. 
\begin{table}[h]
\begin{center}
\begin{tabular}{|cc|ll|ll|}
\hline
& & \multicolumn{2}{c|}{good set}  &  \multicolumn{2}{c|}{all data}  \\
$p$ bin & $p/\mathrm{GeV}$ & $\chi^2/NDF$ & Flux & 
 $\chi^2/NDF$ & Flux \\
\hline   
1 & 11.5 & 19/15 & 9.88$\cdot 10^{-1}$ & 129/20 & 1.00  \\ 
2 & 15.4 & 11/12 & 5.09$\cdot 10^{-1}$ &  40/17 & 5.11$\cdot 10^{-1} $ \\
3 & 20.5 & 20/14 & 2.49$\cdot 10^{-1}$ &  36/21 & 2.46$\cdot 10^{-1} $ \\
4 & 27.4 & 19/16 & 1.19$\cdot 10^{-1}$ &  76/27 & 1.17$\cdot 10^{-1} $ \\
5 & 36.5 & 22/14 & 5.49$\cdot 10^{-2}$ &  85/22 & 5.56$\cdot 10^{-2} $ \\
6 & 48.7 & 12/15 & 2.47$\cdot 10^{-2}$ &  45/31 & 2.47$\cdot 10^{-2} $ \\
7 & 64.9 & 14/11 & 1.08$\cdot 10^{-2}$ &  33/23 & 1.09$\cdot 10^{-2} $ \\
8 & 86.6 & 8/8   & 4.52$\cdot 10^{-3}$ &  48/19 & 4.70$\cdot 10^{-3} $ \\
9 & 115  & 10/6  & 1.86$\cdot 10^{-3}$ &  51/12 & 1.98$\cdot 10^{-3} $ \\
10 & 154 & 2.4/2& 7.78$\cdot 10^{-4} $ &  18/9  & 8.44$\cdot 10^{-4} $ \\
11 & 205 & 0.3/2 & 3.27$\cdot 10^{-4}$ & 10/6 & 3.39$\cdot 10^{-4} $ \\
12 & 274 & 0/0   & 1.05$\cdot 10^{-4} $ & 12/4 & 1.33$\cdot 10^{-4} $ \\
13 & 365 & 0.3/1 & 5.03$\cdot 10^{-5} $ & 0.6/4& 5.18$\cdot 10^{-5} $ \\
14 & 487 & 0/0   & 1.91$\cdot 10^{-5} $ & 1.1/3& 2.00$\cdot 10^{-5} $ \\
15 & 866 &       &         & 0/0  & 6.08$\cdot 10^{-6} $ \\
16 & 1155 & 0/0   & 2.76$\cdot 10^{-5} $ & 0/0  & 2.76$\cdot 10^{-7} $ \\
17 & 1540 & 0/0   & 9.23$\cdot 10^{-7} $ & 3/3  & 9.10$\cdot 10^{-8} $ \\
18 & 2054 &      & &      & \\
\hline
\end{tabular}
\end{center}
\caption{Average flux in $\mathrm{m^{-2} \,  sr^{-1} \,  s^{-1} \,  GeV^{-1}}$}
\label{shrat}
\end{table}
The five datasets discussed above have a large impact on the $\chi^2$ of the relative 
difference in these bins. 
Therefore, we will exclude them when adjusting the shape 
according to the measurements. 
We fit a third degree polynomial to the logarithm of 
the flux as a
function of the logarithm of momentum.
We parameterize this function as follows:
\begin{equation}
\begin{split}
H(y) & = H_1 \cdot (y^3/2-5y^2/2+3y) \\
     &  + H_2 \cdot (-2y^3/3+3y^2-10y/3+1) \\ 
     &  + H_3 \cdot (y^3/6-y^2/2+y/3)\\
     &  + S_2 \cdot (y^3/3-2y^2+11y/3-2) \\
y & = ~^{10}\log(p/{\rm GeV}) \\
F(p) & = 10^{H(y)} \; \mathrm{m^{-2}\, sr^{-1} \, s^{-1} \, GeV^{-1}} 
\end{split}
\end{equation}
This parameterization is similar to the one used by \cite{bugaev}, however
the fit variables are chosen such that they have a simple interpretation: 
$H_1$, $H_2$, and $H_3$ represent the logarithm
of the differential flux at 10, 100 and 1000 GeV,
$S_2$ represents the exponent of the 
differential flux at 100 GeV. 
The $\chi^2/NDF$ of this 
fit is 8/12, the correlation matrix is shown in appendix D.
The fitted parameters are listed below, as well as the 
equivalent values from our 
reference shape. 
\begin{center}
\begin{tabular}{l|cccc}
& $H_1$ & $H_2$ &  $H_3$ & $S_2$ \\ \hline
Reference & 0.127          & -2.539             &  -5.86         & -2.00 \\
Fit &  $ 0.135 \pm 0.002 $ & $-2.529\pm 0.004 $ &  $ -5.76 \pm 0.03 $ & $ -2.10 \pm  0.03 $ \\
\end{tabular}
\end{center}

\begin{figure}[htbp]
\begin{center}
\epsfig{file=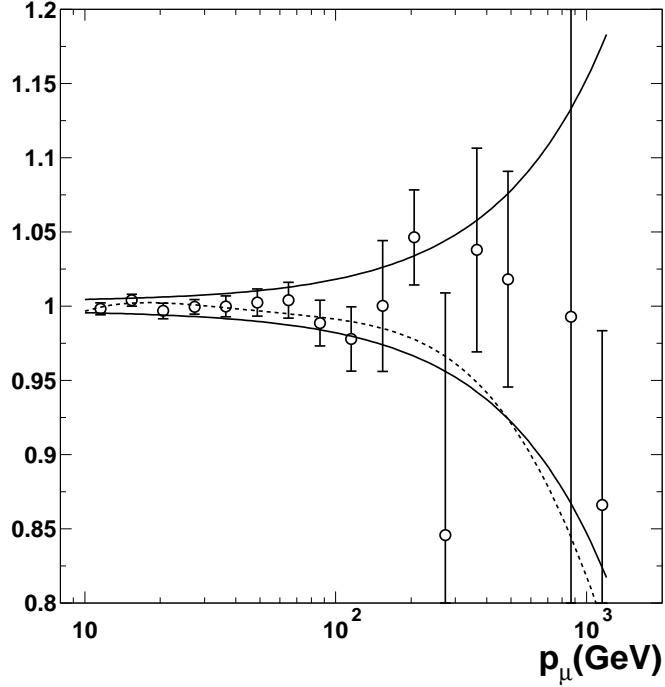,height=10cm,angle=0}
\caption{The ratio between the normalized flux data 
and the fitted function. The band indicates the uncertainty
of the fit. The dashed curve 
represents the normalized  reference shape \cite{bugaev}.}
\label{shnew}
\end{center}
\end{figure}

Figure \ref{shnew} contains a $1 \sigma$ error band, 
which can be approximated by
\begin{eqnarray}
  \delta  =
     0.003 + 0.00015 \cdot \frac{p}{\mathrm{GeV}}
\label{delta}
\end{eqnarray}
It represents the size of the combined relative 
experimental uncertainties as 
a function of momentum bin.

%
%
The shape uncertainty at reference momenta
of 10, 100 and 1000 GeV are 0.5\%, 1.8\% and 15\%
respectively. 
Above 200 GeV the uncertainty rapidly increases, indicating that 
more measurements are needed at these momenta.

\bigskip

\underline{4.2 The absolute normalization of the muon spectrum}

\bigskip

In section 4.1 we fitted a functional shape to all 
renormalized datasets. The renormalization was such that the integral 
flux above 10 GeV corresponds to the calculation by Bugaev. We will now
fit the functional shape obtained in 4.1 to the datasets of those experiments 
providing an absolute flux measurement, while leaving the normalization as 
a free parameter. Therefore we fit the function
\begin{eqnarray}
F(p) & = C \cdot 10^{H(y)}
\end{eqnarray}
The result is shown in table \ref{norm}.
\begin{table}[htbp]
\begin{center}
\begin{tabular}{|l|c|c|}
\hline
Data set & $\rm \chi^2/NDF$ & Fitted Normalization $C$ \\
\hline
 Allkofer 1971 \cite{allkofer1971} & 117/8  & $1.043 \pm 0.006$ \\ 
 Ayre 1975 \cite{ayre1975}         & 286/44 & $0.964 \pm 0.002$ \\ 
 Bateman 1971 \cite{bateman1971}   & 8/8    & $0.860 \pm 0.008$ \\
 Green 1979 \cite{green1979}       & 2.4/4  & $0.967 \pm 0.022$ \\ 
 Tsuji  1998 \cite{tsuji1998}      & 16/13  & $0.948 \pm 0.014$ \\
 De Pascale 1993 \cite{depascale1993} 
                                   & 7/5    & $0.787 \pm 0.015$ \\
 Kremer 1994 data \cite{kremer1999}& 9/6    & $0.811 \pm 0.009$ \\ 
 Kremer 1997 data \cite{kremer1999}& 11/6   & $0.820 \pm 0.008$ \\ 
\hline
\end{tabular}
\end{center}
\caption{Normalization factor $C$ with respect to the integrated flux as calculated by Bugaev}
\label{norm}
\end{table}
We again ignore the datasets with a very high $\chi^2/NDF$ (Allkofer, Ayre).
The three remaining data sets with the largest 
normalization factors (Bateman, Green and Tsuji) 
are measurements 
performed with solid iron magnet spectrometers, 
whereas the other three 
(Kremer 1994 and 1997 and De Pascale) use the same superconducting magnet. \\
We will first average the normalizations performed by the same collaboration
(Bateman and Green, Kremer), and afterwards calculate the normalization 
measurements performed by the solid iron magnet spectrometers and the superconducting 
magnet spectrometers, which gives the following results: 

\begin{center}
\begin{tabular}{l|l}
Solid iron magnets & $0.937 \pm 0.012 $\\
Superconducting magnet& $0.811\pm 0.007$ \\ 
\end{tabular}
\end{center}
%
These values are clearly not in agreement. 
We have no explanation for this observation.
We will simply 
take the average of these two 
values to be our normalization 
and half the difference to be the uncertainty. We 
arrive to our final value of $0.874 \pm 0.063$, 
thus a normalization with a relative uncertainty of 7 \%. 

\bigskip

\underline{4.3 The muon spectrum}

\bigskip

In the preceding sections we have parameterized the muon spectrum at 
sea-level. We summarize the parameters obtained:

\begin{center}
\begin{tabular}{ccccc}
$C$ & $H_1$ & $H_2$ &  $H_3$ & $S_2$\\ 
\hline 
 $0.874 \pm 0.063$ &  $ 0.135 \pm 0.002 $ &  $ -2.529 \pm 0.004 $ 
   & $ -5.76 \pm  0.03 $ & $-2.10 \pm 0.03$ \\
\end{tabular}
\end{center} 

\begin{figure}[htbp]
\begin{center}
\epsfig{file=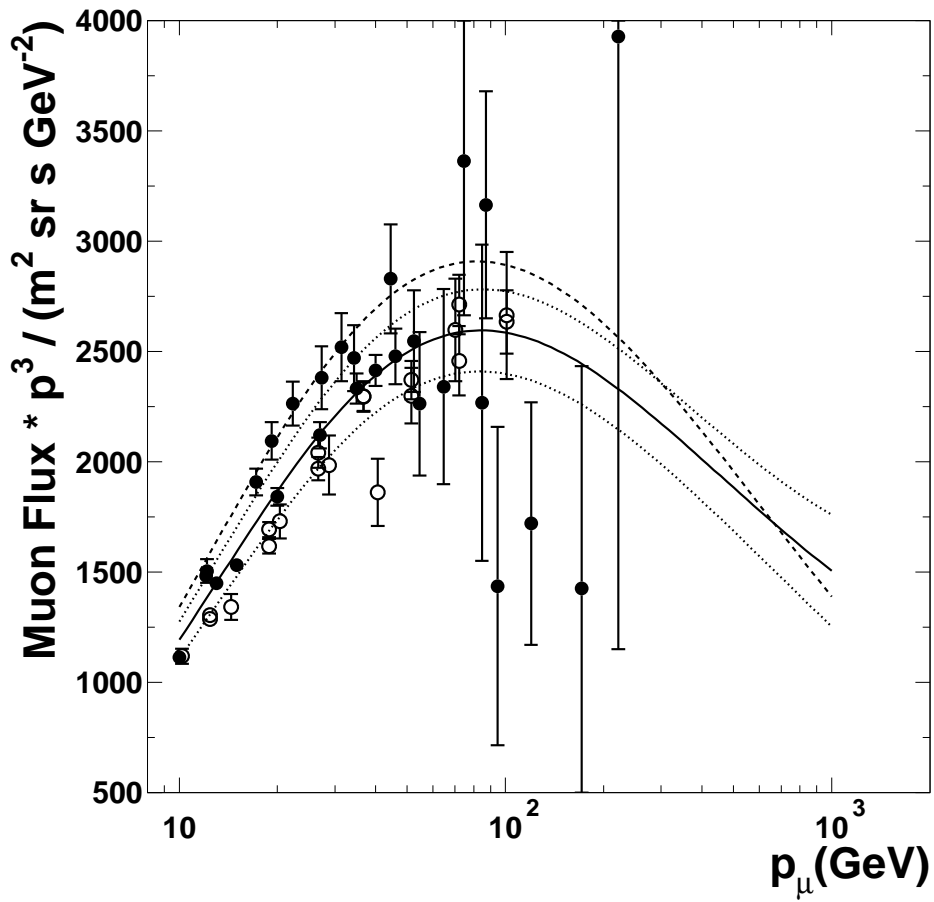,height=13cm,angle=0}
\caption{The result on the muon flux. 
The dotted lines show the 1 sigma error band, whereas 
the dashed curve is the description by Bugaev. 
The points are the data used in the 
normalization procedure.
The open points stand for experiments using a superconducting magnet, 
the black points indicate conventional magnets. 
}
\label{shfinal}
\end{center}
\end{figure}

Our description, as well as the calculation from Bugaev, and the measurements 
used for the normalization  are shown in figure \ref{shfinal}.
The error is given be the estimated normalization uncertainty of $7\%$
and the shape error in (\ref{delta}), added in quadrature. 
\newpage

If we compare our description of the differential flux to the 
theoretical description
of Bugaev \cite{bugaev} we get the following :
\begin{center}
\begin{tabular}{lcccl}
Momentum:        & 10 GeV & 100 GeV &  1000 GeV &\\ 
\hline
Our Result & $1.19 \pm 0.08$   & $(2.59 \pm 0.19) \cdot 10^{-3}$ 
&  $ (1.52\pm 0.26) \cdot 10^{-6}$ 
& $\rm m^{-2} \, sr^{-1} \, s^{-1} \, GeV^{-1}$ \\
Bugaev et al.     & $1.34$   & $2.89\cdot 10^{-3}$ &  
$ 1.39\cdot 10^{-6}$ &  $\rm m^{-2} \, sr^{-1} \, s^{-1} \, GeV^{-1}$ \\
\end{tabular}
\end{center}
At 10 GeV the measured flux is 
89\% of the calculation of Bugaev. However, the measured  
shape is slightly less steep, and at 1 TeV 
we arrive to a value which is close to the predicted one.


\bigskip


\underline{5. Charge Ratio}

\bigskip

The charge ratio $R_{\mu}$ is defined as the ratio of 
vertical fluxes for positive and negative muons at sea level.

The measured charge ratios together with the published
uncertainties are listed in appendix C.
Figure~\ref{allall} shows all values as a function of momentum.
\begin{figure}[htbp]
\begin{center}
\mbox{\epsfig{file=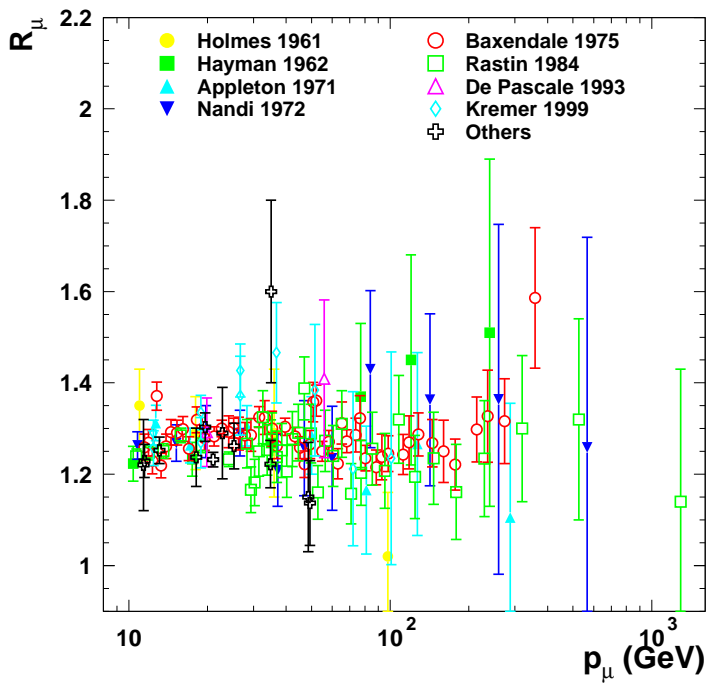,height=10cm,angle=0}}
\caption{Measured charge ratios.}
\label{allall}
\end{center}
\end{figure}

\begin{figure}[htbp]
\begin{center}
\mbox{\epsfig{file=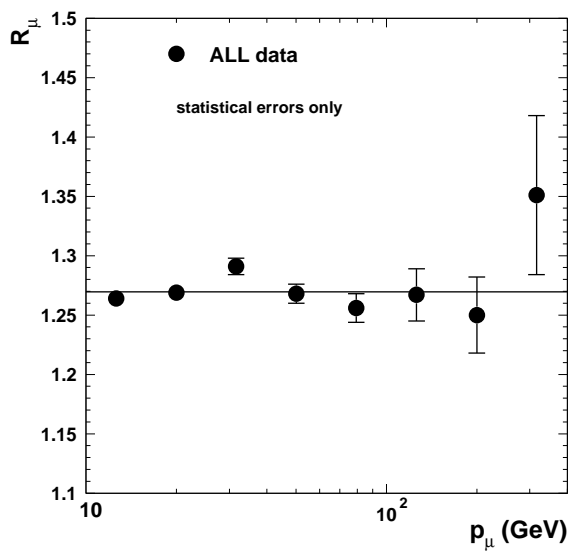,height=8cm,angle=0}}
\caption{Average charge ratio values.}
\label{allstat}
\end{center}
\end{figure}

In order to study the momentum dependence we 
have grouped all 15 data sets into momentum bins 
chosen to be equidistant in $\log p$.
We have combined the different measurements by assuming
that they are uncorrelated.
The bin size is relatively large, since a
strong momentum dependence is not expected.
The result is shown in figure~\ref{allstat}.
The two data points around $500 \, \mathrm{GeV}$ and 
the single measurement above $1 \, \mathrm{TeV}$ 
have huge uncertainties ($ \sim 20\%$) and are therefore not included
in the figure.

For all eight momentum bins the $\chi^2$ values are good or at least 
acceptable; this 
implies the various experimental data agree among each other.
\\
The third momentum yields the highest charge ratio. However, 
this result can not be attributed to a single `outlier'.

\medskip

To see if there is a momentum dependence we have 
performed the following fits to the  charge
ratio values shown in figure~\ref{allstat} as a function of
$\log p$:

\medskip

a) $f(\log p) = R_{\mu}^0 = \mathrm{const} $
\\
This gives a good fit with $\chi^2/NDF = 158/142$.
The resulting charge ratio of
\begin{eqnarray}
  R_{\mu}^0 = 1.270 \pm 0.003
\end{eqnarray}
is displayed in figure~\ref{allstat} as horizontal line. 

\medskip

b) $f(\log p) = R_{\mu}^1 + S_{\mu}^1 \cdot \log(p/\mathrm{GeV})$
\\
Naturally this fit is satisfactory, too.
The slope comes out as 
\begin{eqnarray}
  S^1_{\mu} = 0.006 \pm 0.011
\end{eqnarray}
which is compatible with zero.

\medskip

Therefore, the measured charge ratios are consistent
with the hypothesis of being momentum independent in the
range 
$10 \,  \mathrm{GeV} \leq p \leq 300 \, \mathrm{GeV}$.

We have looked at the data in more detail and 
tried to answer the following questions:
\\
i) Do the different experiments agree with each other ?
\\
ii) Is the `peak' at about $30 \, \mathrm{GeV}$ significant ?

\begin{figure}[htbp]
\begin{center}
\mbox{\epsfig{file=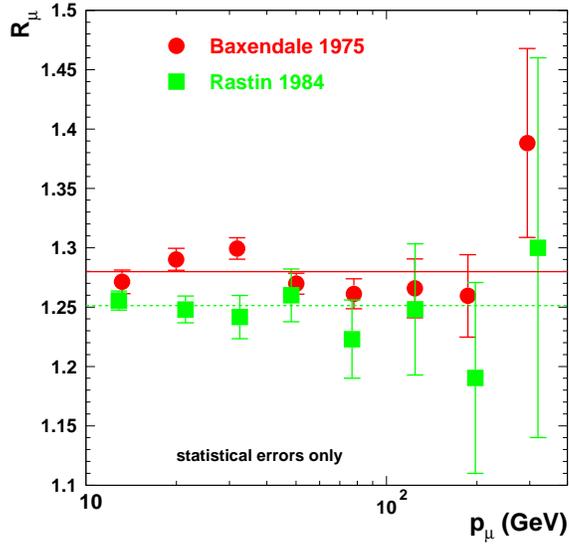,height=8cm,angle=0}}
\caption{Comparison of charge ratios as measured by 
Baxendale et al.\cite{baxendale1975} and Rastin et al.\cite{rastin1984b}.}
\label{baxrast}
\end{center}
\end{figure}

The previous statistical analyses and figure~\ref{allall}
seem to imply the answer `yes' to the 
first question.
However, when separately plotting the two (by far) most precise 
data sets (Baxendale 1975\cite{baxendale1975} and 
Rastin 1984\cite{rastin1984b}), 
one  finds the discrepancy
displayed in figure~\ref{baxrast}.
Averaged over all momenta the 
mean values 
\begin{eqnarray}
 R_{\mu}^0 (Baxendale)  &  = &  1.2799 \pm 0.0042 \\  
 R_{\mu}^0 (Rastin)     & =  & 1.2511 \pm 0.0058
\end{eqnarray}
disagree on the  $ 4 \sigma $ level.
To reduce the discrepancy to about $1 \sigma$ 
we assume - in the spirit of the Particle Data Group\cite{pdg} -
that for {\it all} experiments an additional systematic
error of $\pm 0.015$ must be added,
in form of a scale uncertainty common to all measurements of one 
experiment, independent of momentum.
Clearly, this is a crude model!

\begin{figure}[htbp]
\begin{center}
\mbox{\epsfig{file=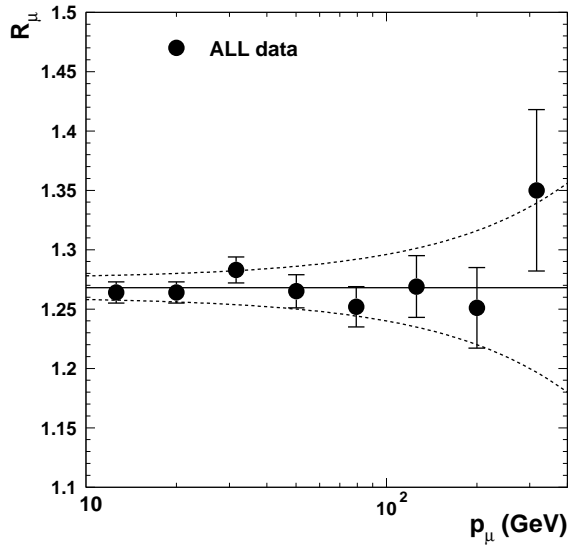,height=8cm,angle=0}}
\caption{Summary of measured charge ratios and a parameterisation of
the central value and the 
uncertainties.}
\label{all}
\end{center}
\end{figure}

%
Including these errors results in the charge ratios displayed in 
figure~\ref{all}. Note that the values are quite close to those
in figure~\ref{allstat}, while the error bars are enlarged.
The corresponding numbers are listed in table~\ref{chratio}.
The momenta are calculated from the 
central values of the logarithmic bins.
\begin{table}
\begin{center}
\begin{tabular}{ccc}
$\log \, p/\mathrm{GeV}$ & $p/\mathrm{GeV}$ & $R_{\mu}$ \\
\hline
1.0-1.2  & 12.6 &  1.264 $\pm$ 0.009  \\
1.2-1.4  & 20.0 &  1.264 $\pm$ 0.009  \\
1.4-1.6  & 31.6 &  1.283 $\pm$ 0.011  \\
1.6-1.8  & 50.1 &  1.265 $\pm$ 0.014  \\
1.8-2.0  & 79.4 &  1.252 $\pm$ 0.017  \\
2.0-2.2  & 126  &  1.269 $\pm$ 0.026  \\
2.2-2.4  & 200  &  1.251 $\pm$ 0.034  \\
2.4-2.6  & 316  &  1.350 $\pm$ 0.068  \\
\end{tabular}
\end{center}
\caption{Average charge ratios}
\label{chratio}
\end{table}

There is no final answer to question ii). 
While the measurements by Baxendale (and also others, with 
larger errors) indicate   
an increase
of the charge ratio at momenta around $30 \, \mathrm{GeV}$, 
the data by Rastin do not support this hypothesis.  
For the moment the measurements are 
consistent with the simple hypothesis of a 
momentum independent charge ratio. 

We try to summarize the charge ratio measurements 
and their uncertainty ($68\%$ CL) with the following formula: 
\begin{eqnarray}
  R_{\mu} = 1.268 \pm 
   \left[  0.008 + 0.0002 \cdot \frac{p}{\mathrm{GeV}}  \right] 
\label{band}
\end{eqnarray}
in the momentum range $10 - 300 \, \mathrm{GeV}$.
Figure~\ref{all} shows the corresponding mean value and the 
error band. 
The central value is the mean  of all measurements,
taking into account the additional systematic error of $0.015$.
The momentum dependent error is estimated such that it is 
roughly of the same size  
as the uncertainties of the corresponding data points
in figure~\ref{all} and table~\ref{chratio}. 

Theoretical models of atmospheric showers must 
be able to reproduce these data, the calculated charge ratios 
should fall into
the band given in equation (\ref{band}) and figure~\ref{all}.
Clearly, at high momenta  precise data is still lacking.
Momenta above a few hundred GeV are of particular interest, 
since a growing influence of kaons 
and a resulting increase of the charge ratio is 
predicted\cite{gaisser}.


\bigskip

\underline{6. Summary and conclusions}

\bigskip

We have combined the published data on the vertical muon spectrum and 
charge ratio at sea-level. 
In this comparison we have found that the differential spectrum can 
be described using a simple 
formula.
The shape of the momentum spectrum is well measured at momenta 
below 100 GeV. Above  
200 GeV only a few data points exist, therefore the uncertainty 
increases to 17 \% at 1 TeV.

Several experiments measure the absolute normalization of the spectrum. 
Our combined result is compared to the calculation 
by Bugaev et al. 
At 10 GeV the measured flux is 11 \% below the calculated one. 

The charge ratio is reported by many experiments. 
The combined result favors a momentum independent 
value of the charge ratio of 1.268. The error on the charge ratio 
increases rapidly above 200 GeV
due to a lack of precise experimental data in that region. 

\bigskip

\underline{7. Acknowledgments}

\bigskip

We thank 
Dieter Heck for  helping us to run CORSIKA and 
Albert van Mil 
for collecting charge ratio data from the literature.

\bigskip


\underline{References}

\bibliography{mucompil}   

\newpage

\underline{APPENDIX A} Measurements.

\bigskip

The following three tables list the experiments/publications 
we considered, 
in chronological order, 
together with 
the most important parameters. 

\bigskip

Some entries are missing since the corresponding figures 
are not published. In particular the zenith angle regime 
is characterized frequently only 
by verbal expressions like `near vertical'. 
We distinguish three magnet types: solid {\it iron} and conventional coil,
{\it air} gap magnet with conventional coil and 
air gap magnet with {\it superconducting} coil.   
The period of data taking is indicated by year and month, e.g.
59/11 stands for November 1959.

\bigskip

The following remarks
refer to the experiment numbers in column 1 of the tables. 
The figures mentioned are those of the corresponding publication. 


\bigskip

\noindent
1) 
Not used, since data set is a subset of 3)\cite{owen1951}.

\noindent
2) 
Spectrum data read off from figure 1;
charge ratio taken from 3)\cite{owen1951}.

\noindent
6) 
Spectrum normalized to Rossi\cite{rossi1948}; 
data read off from figures 8 and 15. 

\noindent
7)
Spectrum normalized to Rossi.

\noindent
8)
Spectrum normalized to Rossi; data read off from figures 1 and 3. 

\noindent
9)
The two values at $p = 240 \, \mathrm{GeV}$ are 
not statistically independent. We have calculated the arithmetic mean 
of the two figures and took the smaller
of the two statistical errors as the uncertainty.

\noindent
10)
We do not use these data, 
which were obtained 
in the equator region, 
where the  
geomagnetic cutoff is large (14 GeV) and may influence the 
muon flux.

\noindent
11)
The spectrum data form a subset of those 
published in 13)\cite{appleton1971}. 
Nevertheless we consider it 
separately, since the normalization procedures are slightly different.
When calculating world averages we exclude these data. 
We do not use the charge ratio data, since they are included in the superset
published in 13)\cite{appleton1971}.

\noindent
13) 
Spectrum: no absolute flux determination, only 
normalization to previous measurements by other experiments. 
Charge ratio: A few measurements are given with slightly asymmetric errors; 
they have been `symmetrized' by shifting the central value
to the center of the error interval.

\noindent
15)
Apparatus under concrete shelter 
of $868 \, \mathrm{g/cm^2}$.

\noindent
16)
We use only the charge ratio value obtained at the town of 
Kiel\cite{allkofer1971}; for the 
other measurements, 
made at muon momenta close to and below 10 GeV in the equator region, 
the  
geomagnetic cutoff is large (14 GeV) and may influence the 
muon flux.

\noindent
20) 
Resolution correction is based upon MDM = 100 GeV;
if MDM of 350 GeV is used, spectrum is flatter and better consistent with 
18) \cite{ayre1975}. Points up
to 100 GeV are considered reliable.

\noindent
21)
Spectrum
data are normalized to an integral intensity at 5 GeV. 
The spectrum provided is the theoretical spectrum that fits the data best.

\noindent
22) 
Above 50 GeV authors question results due to resolution.

\noindent
24, 25) 
Same magnet as 22)\cite{depascale1993}.
Charge ratio and its error read off from figure 2. 

\newpage 

\begin{landscape}

\small

\begin{tabular}{ccccccccccccc}

No & author(s), & name & location & coordinates & alt. &
magnet & detector(s) & period & MDM & zenith & spec. & ratio  \\
& reference  & & &  & /m & & & & /GeV & & &  \\
\hline
 & & & & \\

1 & 
B.G. Owen and &
  &
Manchester & 
53$^0$N,  2$^0$W & 
50  &
air & 
Geiger,  &
&
30  & 
$< 8^0$ & 
 no &
 yes \\
&
J.G. Wilson,
1949\cite{owen1949} & 
 &
(Great Britain) &
&
  &
  & 
flash tubes  &
  & 
  &
  &
  &
\\

 & & & & \\

2 &
D.E. Caro &
 &
Melbourne & 
38$^0$S,  145$^0$E &
50  &
air & 
counters &
  & 
$\approx$ 50 &
 &
 yes &
 yes \\
&
et al,
1950\cite{caro1950} &
 & 
(Australia) &
&
  &
  & 
  &
  & 
  &
  &
  &
\\

 & & & & \\

3 & 
B.G. Owen and &
  &
Manchester & 
53$^0$N,  2$^0$W & 
50  &
air & 
Geiger,  &
&
30  & 
$< 8^0$ & 
no &
 yes \\
&
J.G. Wilson,
1951\cite{owen1951} & 
 &
(Great Britain) &
&
  &
  & 
flash tubes  &
  & 
  &
  &
  &
\\

 & & & & \\

4 & 
I. Filosofo & 
  &
Agordo & 
46$^0$N, 12$^0$E &
600  &
iron & 
counters &
  & 
 & 
 &
 no &
 yes \\
&
et. al, 
1954\cite{filosofo1954} &
 &
(Italy) &
&
  &
  & 
  &
  & 
  &
  &
  &
\\

 & & & & \\

5 & 
B.G. Owen and &
  &
Manchester & 
53$^0$N,  2$^0$W & 
50  &
air & 
counters  &
  & 
 & 
$<10^0$ &
 yes &
 no \\
&
J.G. Wilson,
1955\cite{owen1955} & 
 &
(Great Britain) &
&
  &
  & 
  &
  & 
  &
  &
  &
\\

 & & & & \\

6 & 
J. Pine et al, &
  &
Cornell &
42$^0$N, 76$^0$W &
500  &
air & 
Geiger,  &
  & 
175  & 
 &
 yes &
 yes \\
&
1959\cite{pine1959} &
 &
(USA) &
&
  &
  & 
cloud ch.  &
  & 
  &
  &
  &
\\

 & & & & \\

7 & 
J.E.R. Holmes & 
  &
Manchester & 
53$^0$N,  2$^0$W & 
50  &
air & 
cloud ch.,  &
53-55  & 
 & 
$< 10^0$ &
 yes &
 yes \\
&
et al, 
1961\cite{holmes1961} & 
 &
(Great Britain) &
&
  &
  & 
Geiger  &
  & 
  &
  &
  &
\\

 & & & & \\

8 & 
W. Pak et al, &
  &
Cornell &
42$^0$N, 76$^0$W  &
500  &
air & 
Geiger,  &
  & 
175  & 
&
 yes &
 yes \\
&
1961\cite{pak1961} &
 &
(USA) &
&
  &
  & 
hodosc. &
  & 
  &
  &
  &
\\

 & & & & \\

9 & 
P.J. Hayman and A.W. &
  &
Durham & 
54$^0$N,  1$^0$W & 
70  &
air & 
Geiger,  &
59/11-60/03  & 
657 & 
 &
 yes &
 yes \\
&
Wolfendale, 1962\cite{hayman1962s,hayman1962} &
 &
(Great Britain) &
&
  &
 &
flash tubes  & 
  & 
  &
  &
  &
\\

 & & & & \\

10 & 
O.C. Allkofer &
  &
near  & 
29$^0$W, 0$^0$N &
0 &
air & 
spark ch.  &
  & 
 & 
 &
yes  &
yes  \\
&
et al, 1968\cite{allkofer1968} &
 &
equator  &
22$^0$W, 1$^0$S & 
&
  & 
  &
  & 
  &
  &
  &
\\

 & & & & \\

\hline

\end{tabular}

\begin{tabular}{ccccccccccccc}

No & author(s), & name & location & coordinates & alt. &
magnet & detector(s) & period & MDM & zenith & spec. & ratio  \\
& reference  & & &  & /m & & & & /GeV & & &  \\
\hline

 & & & & \\

11 & 
S.R. Baber &
  &
Nottingham & 
53$^0$N,  1$^0$W & 
52  &
iron & 
Geiger &
64/06-65/05  & 
360  & 
 &
 yes &
 yes \\
&
et al, 1968\cite{baber1968a,baber1968b} &
 &
(Great Britain) &
&
  &
  & 
 &
  & 
  &
  &
  &
\\

 & & & & \\

12 &
A.M. Aurela and A.W. &
 &
Durham &
54$^0$N, 1$^0$W &
70 &
air, &
Geiger, &
64/06-65/01 &
 &
 &
yes &
no \\
&
Wolfendale,
1967\cite{aurela1967} &
 &
(Great Britain) &
&
&
iron &
flash  & 
 &
  & 
  &
  & 
\\

 & & & & \\

13 & 
I.C. Appleton  &
  &
Nottingham & 
53$^0$N,  1$^0$W & 
52  &
iron & 
flash  &
64-68  & 
360  & 
 &
 yes &
 yes \\
&
et al, 1971\cite{appleton1971} &
 &
(Great Britain) &
&
  &
  & 
tubes  &
  & 
  &
  &
  &
\\

 & & & & \\

14 & 
B.J. Bateman &
AMH  &
College Station &
31$^0$N,  96$^0$W & 
80  &
iron & 
spark ch.,  &
& 
& 
 &
 yes &
 no \\
&
et al, 1971\cite{bateman1971} &
 &
(USA)
&
  &
  & 
  &
scint. &
  &
  &
  &
\\

 & & & & \\

15 & 
O.C. Allkofer &
  &
Kiel  & 
54$^0$N,  11$^0$E  &
10 &
iron & 
spark ch.,  &
& 
$> 1000$  & 
 &
yes  &
no  \\
&
et al, 1971\cite{allkofer1971} &
 &
(Germany)  &
&
&
  & 
scint.  &
  & 
  &
  &
  &
\\

 & & & & \\

16 & 
O.C. Allkofer and &
  &
Kiel  & 
54$^0$N,  11$^0$E  &
10 &
iron & 
spark ch.,  &
 & 
 & 
 &
no  &
yes  \\
&
W.D. Dau, 1972\cite{allkofer1972} &
 &
(and equator)  &
&
&
  & 
scint.  &
  & 
  &
  &
  &
\\

 & & & & \\

17 & 
B.C. Nandi &
  &
Durgapur  & 
24$^0$N,  87$^0$E  &
70 &
iron & 
flash tubes,  &
69/02-70/02 &
$ 985 $  & 
 &
yes  &
yes  \\
&
et al, 1972\cite{nandi1972, nandi1972s} &
 &
(India)  &
&
&
  & 
Geiger  &
  & 
  &
  &
  &
\\

 & & & & \\

18 & 
C.A. Ayre &
MARS  &
Durham & 
54$^0$N,  1$^0$W & 
70  &
iron & 
scint.,  &
72/05-73/01 &
670 & 
 &
 yes &
 no \\
&
et al, 1975\cite{ayre1975} &
 &
(Great Britain) &
&
  &
 &
flash tubes  & 
  & 
  &
  &
  &
\\

 & & & & \\

19 & 
J.M. Baxendale &
  &
Durham & 
54$^0$N,  1$^0$W & 
70  &
iron & 
scint.,  &
72/02-72/12 &
 & 
 &
 yes &
 yes \\
&
et al, 1975\cite{baxendale1975} &
 &
(Great Britain) &
&
  &
 &
flash tubes  & 
  & 
  &
  &
  &
\\

 & & & & \\
20 & 
P.J. Green &
AMH  &
Houston &
30$^0$N,  95$^0$W & 
10  &
iron & 
spark ch.,  &
&
345 & 
$< 9^0$ & 
 yes &
 no \\
&
et al, 1979\cite{green1979} &
 &
(USA)
&
  &
  & 
&
scint. &
  &
  &
  &
\\

 & & & & \\

\hline

\end{tabular}

\begin{tabular}{ccccccccccccc}

No & author(s), & name & location & coordinates & alt. &
magnet & detector(s) & period & MDM & zenith & spec. & ratio  \\
& reference  & & &  & /m & & & & /GeV & & &  \\
\hline

 & & & & \\

21 & 
B.C. Rastin &
  &
Nottingham & 
53$^0$N,  1$^0$W & 
52  &
iron & 
flash tubes,  &
74/09-78/05 &
3400  & 
 &
 yes &
 yes \\
&
1984\cite{rastin1984a,rastin1984b} &
 &
(Great Britain) &
&
  &
  & 
scint.  & 
  &
&
  &
  &
\\

 & & & & \\

22 & 
M.P. De Pascale &
MASS &
Prince Albert &
53$^0$N, 106$^0$W &
600 &
air &
MWPC, &
89/08 &
118 &
 &
yes &
yes \\
&
et al,  
1993\cite{depascale1993} &
 &
(Canada) &
&
 &
(superc.)  & 
scint., TOF &
  & 
  &
  &
  &
\\

 & & & & \\

23 &
S. Tsuji &
 &
Okayama &
34$^0$N, 134$^0$E &
5 & 
iron &
drift, &
92/09-97/12 &
270 &
0$^0$-1$^0$  &
yes &
no \\
&
et al, 
1998\cite{tsuji1998} & 
 &
(Japan) &
&
 &
  & 
scint.  &
  & 
  &
  &
  &
\\

 & & & & \\

24 &
J. Kremer 
&
CAPRICE &
Lynn Lake &
57$^0$N, 101$^0$W  &
360 &
air &
prop., drift, &
94/07 &
175  &
0$^0$-20$^0$ &
yes &
yes\\
&
et al, 
1999\cite{kremer1999} & 
 &
(Canada) &
&
&
(superc.) &
TOF, RICH 
  & 
  &
  &
  &
\\

 & & & & \\

25 &
J. Kremer & 
CAPRICE &
Fort Sumner & 
34$^0$N, 104$^0$W & 
1270 &
air &
drift, &
97/04-97/05 &
175  &
0$^0$-20$^0$ &
yes &
yes\\
&
et al, 
1999\cite{kremer1999}  & 
 &
(USA) &
&
&
(superc.) &
TOF, RICH  &
  & 
  &
  &
  &
\\

\hline

\end{tabular}

\end{landscape}

\newpage 

\begin{multicols}{2}

\underline{APPENDIX B} Muon flux data.

\medskip

The following lists contain all 
spectrum
data for momenta above $10 \, \mathrm{GeV}$. 
The flux is given in  ($\rm m^{-2} \, sr^{-1} \, s^{-1} \, GeV$).
Each line contains the momentum 
in $\mathrm{GeV}$ together with the 
published value and uncertainty of
the flux. 
Altitude corrections have been applied later and are not
included in the figures 
listed here.

\begin{itemize}
\item D.E. Caro et al, 1950\cite{caro1950}

\begin{center}
\begin{tabular}{cc}
  12.2  &  (6.46 $\pm$ 0.75)$\cdot 10^{-1}$ \\
  16.9  &  (2.59 $\pm$ 0.26)$\cdot 10^{-1}$ \\
  17.8  &  (2.38 $\pm$ 0.25)$\cdot 10^{-1}$ \\
  26.5  &  (7.4  $\pm$ 0.9)$\cdot 10^{-2}$ \\
  51    &  (1.3  $\pm$ 0.2)$\cdot 10^{-2}$ \\
\end{tabular}
\end{center}

\item B.G. Owen and J.G. Wilson, 1955\cite{owen1955}

\begin{center}
\begin{tabular}{cc}
  10.0 & 1.09 $\pm$ 0.03 \\
  15.0 & (4.36 $\pm$ 0.11)$\cdot 10^{-1}$ \\             
  20.0 &  (2.20 $\pm$ 0.06)$\cdot 10^{-1}$ \\
\end{tabular}
\end{center}

\item J.Pine et al, 1959\cite{pine1959}

\begin{center}
\begin{tabular}{cc}

  11.6  & (8.26 $\pm$ 0.99)$\cdot 10^{-1}$ \\   
  13.7  & (5.24 $\pm$ 0.84)$\cdot 10^{-1}$ \\
  16.3  & (3.05 $\pm$ 0.49)$\cdot 10^{-1}$ \\
  18.1  & (2.79 $\pm$ 0.33)$\cdot 10^{-1}$ \\
  19.8  & (1.91 $\pm$ 0.31)$\cdot 10^{-1}$ \\
  25.5  & (1.16 $\pm$ 0.19)$\cdot 10^{-1}$ \\
  29.8  & (7.68 $\pm$ 0.92)$\cdot 10^{-2}$ \\
  35.4  & (4.95 $\pm$ 0.79)$\cdot 10^{-2}$ \\
  46.0  & (1.99 $\pm$ 0.32)$\cdot 10^{-2}$ \\
  59.1  & (9.29 $\pm$ 1.49)$\cdot 10^{-3}$ \\
  83.2  & (4.60 $\pm$ 0.87)$\cdot 10^{-3}$ \\
 125.   & (1.34 $\pm$ 0.33)$\cdot 10^{-3}$ \\

\end{tabular}
\end{center}

\item W.Pak et al, 1961\cite{pak1961}

\begin{center}
\begin{tabular}{cc}

  12.9  & (6.14 $\pm$ 0.68)$\cdot 10^{-1}$ \\   
  15.8  & (3.35 $\pm$ 0.37)$\cdot 10^{-1}$ \\
  19.8  & (1.89 $\pm$ 0.21)$\cdot 10^{-1}$ \\
  24.1  & (1.35 $\pm$ 0.19)$\cdot 10^{-1}$ \\ 
  26.4  & (1.03 $\pm$ 0.11)$\cdot 10^{-1}$ \\
  37.4  & (3.77 $\pm$ 0.41)$\cdot 10^{-2}$ \\
  58.9  & (1.29 $\pm$ 0.14)$\cdot 10^{-2}$ \\  

\end{tabular}
\end{center}


\item J.E.R. Holmes et al, 1961\cite{holmes1961}

\begin{center}
\begin{tabular}{cc}

  11  & (7.88 $\pm$ 0.32)$\cdot 10^{-1}$ \\     
  13  & (5.90 $\pm$ 0.30)$\cdot 10^{-1}$ \\       
  16  & (3.68 $\pm$ 0.18)$\cdot 10^{-1}$ \\
  19  & (2.33 $\pm$ 0.12)$\cdot 10^{-1}$ \\
  23  & (1.67 $\pm$ 0.12)$\cdot 10^{-1}$ \\
  28  & (1.04 $\pm$ 0.07)$\cdot 10^{-1}$ \\
  36  & (5.75 $\pm$ 0.40)$\cdot 10^{-2}$ \\
  49  & (2.02 $\pm$ 0.20)$\cdot 10^{-2}$ \\
  67  & (8.65 $\pm$ 1.30)$\cdot 10^{-3}$ \\
  89  & (4.38 $\pm$ 0.70)$\cdot 10^{-3}$ \\
 134  & (1.42 $\pm$ 0.26)$\cdot 10^{-3}$ \\
 271  & (2.8 $\pm$ 0.6)$\cdot 10^{-4}$ \\
1160  & (4.8 $\pm$ 2.3)$\cdot 10^{-6}$ \\

\end{tabular}
\end{center}

\item P.J. Hayman and A.W. Wolfendale 1962\cite{hayman1962s}

\begin{center}
\begin{tabular}{cc}
 
 10.8  & (8.51 $\pm$ 0.26)$\cdot 10^{-1}$ \\
 12.4  & (6.14 $\pm$ 0.17)$\cdot 10^{-1}$ \\  
 14.6  & (4.35 $\pm$ 0.11)$\cdot 10^{-1}$ \\
 17.8  & (2.52 $\pm$ 0.07)$\cdot 10^{-1}$ \\
 22.6  & (1.39 $\pm$ 0.04)$\cdot 10^{-1}$ \\
 31.3  & (5.85 $\pm$ 0.15)$\cdot 10^{-2}$ \\
 42.3  & (2.88 $\pm$ 0.11)$\cdot 10^{-2}$ \\
 56.1  & (1.22 $\pm$ 0.05)$\cdot 10^{-2}$ \\
 72.5  & (5.75 $\pm$ 0.46)$\cdot 10^{-3}$ \\
 88.1  & (3.27 $\pm$ 0.29)$\cdot 10^{-3}$ \\
 112   & (1.36 $\pm$ 0.14)$\cdot 10^{-3}$ \\
 153   & (5.18 $\pm$ 0.93)$\cdot 10^{-4}$ \\
 244   & (1.14 $\pm$ 0.22)$\cdot 10^{-4}$ \\
 413   & (1.98 $\pm$ 0.99)$\cdot 10^{-5}$ \\
 894   & (1.84 $\pm$ 1.01)$\cdot 10^{-6}$ \\

\end{tabular}
\end{center}

\item A.M. Aurela and A.W. Wolfendale 1967\cite{aurela1967}

\begin{center}
\begin{tabular}{cc}
 
 15.1 & (4.25 $\pm$ 0.16)$\cdot 10^{-1}$ \\
 41.5 & (3.40 $\pm$ 0.44)$\cdot 10^{-2}$ \\
 82.1 & (4.10 $\pm$ 0.35)$\cdot 10^{-3}$ \\

\end{tabular}
\end{center}

%
%
%

\vspace*{2.5cm}



\item S.R. Baber et al, 1968\cite{baber1968a}

\begin{center}
\begin{tabular}{cc}

  11.60  & (7.77 $\pm$ 0.26)$\cdot 10^{-1}$ \\    
  15.22  & (4.22 $\pm$ 0.21)$\cdot 10^{-1}$ \\ 
  19.20  & (2.42 $\pm$ 0.12)$\cdot 10^{-1}$ \\    
  24.00  & (1.39 $\pm$ 0.07)$\cdot 10^{-1}$ \\     
  33.5   & (5.78 $\pm$ 0.35)$\cdot 10^{-2}$ \\ 
  50.0   & (1.90 $\pm$ 0.16)$\cdot 10^{-2}$ \\ 
  81.0   & (4.59 $\pm$ 0.60)$\cdot 10^{-3}$ \\ 
 127.0   & (1.14 $\pm$ 0.18)$\cdot 10^{-3}$ \\ 
 266.0   & (1.00 $\pm$ 0.24)$\cdot 10^{-4}$ \\ 
 810.0   & (2.11 $\pm$ 0.55)$\cdot 10^{-6}$ \\ 

\end{tabular}
\end{center}

\item I.C. Appleton et al, 1971\cite{appleton1971}

\begin{center}
\begin{tabular}{cc}

  12.84  & (2.92 $\pm$ 0.04)$\cdot 10^{-1}$ \\      
  17.2   & (1.46 $\pm$ 0.02)$\cdot 10^{-1}$ \\
  24.3   & (5.78 $\pm$ 0.13)$\cdot 10^{-2}$ \\
  33     & (2.54 $\pm$ 0.06)$\cdot 10^{-2}$ \\
  43.4   & (1.09 $\pm$ 0.10)$\cdot 10^{-2}$ \\
  45.6   & (1.07 $\pm$ 0.09)$\cdot 10^{-2}$ \\
  48.2   & (9.00 $\pm$ 0.81)$\cdot 10^{-3}$ \\
  51.0   & (6.95 $\pm$ 0.67)$\cdot 10^{-3}$ \\
  54.2   & (6.84 $\pm$ 0.63)$\cdot 10^{-3}$ \\
  57.9   & (5.20 $\pm$ 0.52)$\cdot 10^{-3}$ \\
  62.0   & (4.79 $\pm$ 0.47)$\cdot 10^{-3}$ \\
  66.9   & (2.80 $\pm$ 0.33)$\cdot 10^{-3}$ \\
  72.6   & (2.76 $\pm$ 0.31)$\cdot 10^{-3}$ \\
  79.3   & (2.02 $\pm$ 0.23)$\cdot 10^{-3}$ \\
  87.5   & (1.57 $\pm$ 0.17)$\cdot 10^{-3}$ \\
  97.6   & (1.17 $\pm$ 0.13)$\cdot 10^{-3}$ \\
 110.3   & (6.97 $\pm$ 0.94)$\cdot 10^{-4}$ \\
 129.9   & (3.00 $\pm$ 0.54)$\cdot 10^{-4}$ \\
 149.4   & (2.93 $\pm$ 0.29)$\cdot 10^{-4}$ \\
 181.5   & (1.34 $\pm$ 0.24)$\cdot 10^{-4}$ \\
 230.9   & (6.61 $\pm$ 1.41)$\cdot 10^{-5}$ \\
 316.1   & (2.72 $\pm$ 0.59)$\cdot 10^{-5}$ \\
 491.5   & (7.67 $\pm$ 1.69)$\cdot 10^{-6}$ \\
1000.0   & (5.21 $\pm$ 1.34)$\cdot 10^{-7}$ \\

\end{tabular}
\end{center}

\vspace*{4cm}



\item B.J. Bateman et al, 1971\cite{bateman1971}

\begin{center}
\begin{tabular}{cc}

  10.0  &  1.12 $\pm$ 0.03 \\
  13.0  & (6.63 $\pm$ 0.13)$\cdot 10^{-1}$ \\ 
  15.0  & (4.56 $\pm$ 0.09)$\cdot 10^{-1}$ \\
  20.0  & (2.31 $\pm$ 0.05)$\cdot 10^{-1}$ \\
  27.0  & (1.08 $\pm$ 0.03)$\cdot 10^{-1}$ \\
  35.0  & (5.45 $\pm$ 0.16)$\cdot 10^{-2}$ \\ 
  40.0  & (3.78 $\pm$ 0.11)$\cdot 10^{-2}$ \\
  46.0  & (2.55 $\pm$ 0.13)$\cdot 10^{-2}$ \\
  53.0  & (1.70 $\pm$ 0.08)$\cdot 10^{-2}$ \\                       

\end{tabular}
\end{center}

\item O.C. Allkofer et al, 1971\cite{allkofer1971}

\begin{center}
\begin{tabular}{cc}

  11.4  & 1.13 $\pm$ 0.01 \\
  14.8  & (6.04 $\pm$ 0.08)$\cdot 10^{-1}$ \\
  20.5  & (2.51 $\pm$ 0.03)$\cdot 10^{-1}$ \\
  31.4  & (8.01 $\pm$ 0.13)$\cdot 10^{-2}$ \\
  52.3  & (1.89 $\pm$ 0.05)$\cdot 10^{-2}$ \\
  93.0  & (3.38 $\pm$ 0.14)$\cdot 10^{-3}$ \\
 175.0  & (5.19 $\pm$ 0.37)$\cdot 10^{-4}$ \\
 329.0  & (7.84 $\pm$ 1.12)$\cdot 10^{-5}$ \\
 642.0  & (6.40 $\pm$ 1.92)$\cdot 10^{-6}$ \\

\end{tabular}
\end{center}

\item B.C. Nandi and M.S. Sinha, 1972\cite{nandi1972s}

\begin{center}
\begin{tabular}{cc}

  11.8  & (9.43 $\pm$ 0.15)$\cdot 10^{-1}$ \\
  14.0  & (6.38 $\pm$ 0.14)$\cdot 10^{-1}$ \\
  16.4  & (4.21 $\pm$ 0.09)$\cdot 10^{-1}$ \\
  19.7  & (2.72 $\pm$ 0.06)$\cdot 10^{-1}$ \\
  24.2  & (1.41 $\pm$ 0.04)$\cdot 10^{-1}$ \\
  29.6  & (1.01 $\pm$ 0.03)$\cdot 10^{-1}$ \\
  37.1  & (5.40 $\pm$ 0.17)$\cdot 10^{-2}$ \\
  46.9  & (2.99 $\pm$ 0.13)$\cdot 10^{-2}$ \\
  60.0  & (1.45 $\pm$ 0.07)$\cdot 10^{-2}$ \\
  84.0  & (5.58 $\pm$ 0.28)$\cdot 10^{-3}$ \\
 118    & (2.04 $\pm$ 0.17)$\cdot 10^{-3}$ \\
 167    & (6.09 $\pm$ 0.61)$\cdot 10^{-4}$ \\
 260    & (1.96 $\pm$ 0.25)$\cdot 10^{-4}$ \\
 467    & (2.69 $\pm$ 0.67)$\cdot 10^{-5}$ \\
1109    & (1.03 $\pm$ 0.36)$\cdot 10^{-6}$ \\

\end{tabular}
\end{center}

\vfill

\break

\item C.A. Ayre et al, 1975\cite{ayre1975}

\begin{center}
\begin{tabular}{cc}

  21.3  & (2.096 $\pm$ 0.029)$\cdot 10^{-1}$ \\
  22.1  & (1.909 $\pm$ 0.027)$\cdot 10^{-1}$ \\
  23.1  & (1.708 $\pm$ 0.024)$\cdot 10^{-1}$ \\
  24.1  & (1.574 $\pm$ 0.022)$\cdot 10^{-1}$ \\
  25.1  & (1.432 $\pm$ 0.020)$\cdot 10^{-1}$ \\
  26.3  & (1.224 $\pm$ 0.017)$\cdot 10^{-1}$ \\
  27.7  & (1.067 $\pm$ 0.015)$\cdot 10^{-1}$ \\
  29.3  & (9.130 $\pm$ 0.128)$\cdot 10^{-2}$ \\
  31.0  & (7.968 $\pm$ 0.112)$\cdot 10^{-2}$ \\
  33.1  & (6.947 $\pm$ 0.097)$\cdot 10^{-2}$ \\
  35.3  & (5.704 $\pm$ 0.080)$\cdot 10^{-2}$ \\
  38.3  & (4.547 $\pm$ 0.068)$\cdot 10^{-2}$ \\
  40.8  & (4.208 $\pm$ 0.046)$\cdot 10^{-2}$ \\
  41.7  & (3.663 $\pm$ 0.055)$\cdot 10^{-2}$ \\
  42.8  & (3.420 $\pm$ 0.041)$\cdot 10^{-2}$ \\
  44.8  & (2.962 $\pm$ 0.036)$\cdot 10^{-2}$ \\
  45.8  & (2.797 $\pm$ 0.042)$\cdot 10^{-2}$ \\
  47.1  & (2.628 $\pm$ 0.032)$\cdot 10^{-2}$ \\
  49.3  & (2.217 $\pm$ 0.027)$\cdot 10^{-2}$ \\
  50.7  & (2.086 $\pm$ 0.033)$\cdot 10^{-2}$ \\
  52.1  & (2.014 $\pm$ 0.024)$\cdot 10^{-2}$ \\
  55.2  & (1.646 $\pm$ 0.021)$\cdot 10^{-2}$ \\
  57.0  & (1.525 $\pm$ 0.024)$\cdot 10^{-2}$ \\
  58.9  & (1.434 $\pm$ 0.019)$\cdot 10^{-2}$ \\
  63.0  & (1.123 $\pm$ 0.015)$\cdot 10^{-2}$ \\
  65.3  & (1.023 $\pm$ 0.017)$\cdot 10^{-2}$ \\
  67.9  & (9.216 $\pm$ 0.129)$\cdot 10^{-3}$ \\
  73.7  & (7.084 $\pm$ 0.099)$\cdot 10^{-3}$ \\
  76.6  & (6.585 $\pm$ 0.118)$\cdot 10^{-3}$ \\
  80.0  & (5.753 $\pm$ 0.081)$\cdot 10^{-3}$ \\
  88.3  & (4.149 $\pm$ 0.062)$\cdot 10^{-3}$ \\
  93.0  & (3.616 $\pm$ 0.072)$\cdot 10^{-3}$ \\
  98.3  & (3.252 $\pm$ 0.052)$\cdot 10^{-3}$ \\
 112.0  & (2.037 $\pm$ 0.035)$\cdot 10^{-3}$ \\
 118.0  & (1.842 $\pm$ 0.042)$\cdot 10^{-3}$ \\
 128.0  & (1.454 $\pm$ 0.026)$\cdot 10^{-3}$ \\
 145.0  & (9.603 $\pm$ 0.192)$\cdot 10^{-4}$ \\
 160.0  & (7.459 $\pm$ 0.201)$\cdot 10^{-4}$ \\
 177.0  & (5.352 $\pm$ 0.123)$\cdot 10^{-4}$ \\
 214.0  & (2.893 $\pm$ 0.078)$\cdot 10^{-4}$ \\
 236.0  & (2.171 $\pm$ 0.083)$\cdot 10^{-4}$ \\
 274.0  & (1.260 $\pm$ 0.044)$\cdot 10^{-4}$ \\
 358.0  & (5.328 $\pm$ 0.250)$\cdot 10^{-5}$ \\
 367.0  & (4.843 $\pm$ 0.349)$\cdot 10^{-5}$ \\
 442.0  & (2.764 $\pm$ 0.221)$\cdot 10^{-5}$ \\

\end{tabular}
\end{center}

\item P.J. Green et al, 1979\cite{green1979}

\begin{center}
\begin{tabular}{cc}

  12.18  & (8.33 $\pm$ 0.30)$\cdot 10^{-1}$ \\
  19.20  & (2.96 $\pm$ 0.12)$\cdot 10^{-1}$ \\
  31.40  & (8.14 $\pm$ 0.50)$\cdot 10^{-2}$ \\
  52.40  & (1.77 $\pm$ 0.16)$\cdot 10^{-2}$ \\
  87.10  & (4.79 $\pm$ 0.78)$\cdot 10^{-3}$ \\
( 249.90  & (3.95 $\pm$ 0.53)$\cdot 10^{-4}$ ) \\  

\end{tabular}
\end{center}


\item B.C. Rastin, 1984\cite{rastin1984b}

\begin{center}
\begin{tabular}{cc}

  10.69  & 1.156 $\pm$ 0.008 \\
  11.94  & (9.05 $\pm$ 0.06)$\cdot 10^{-1}$ \\
  13.58  & (6.72 $\pm$ 0.04)$\cdot 10^{-1}$ \\
  15.81  & (4.70 $\pm$ 0.03)$\cdot 10^{-1}$ \\ 
  19.05  & (2.97 $\pm$ 0.02)$\cdot 10^{-1}$ \\
  24.14  & (1.63 $\pm$ 0.01)$\cdot 10^{-1}$ \\
  28.35  & (1.03 $\pm$ 0.02)$\cdot 10^{-1}$ \\        
  29.30  & (9.3 $\pm$ 0.2)$\cdot 10^{-2}$ \\
  30.32  & (8.5 $\pm$ 0.2)$\cdot 10^{-2}$ \\
  31.42  & (7.8 $\pm$ 0.2)$\cdot 10^{-2}$ \\
  32.60  & (7.1 $\pm$ 0.2)$\cdot 10^{-2}$ \\
  33.88  & (6.3 $\pm$ 0.1)$\cdot 10^{-2}$ \\
  35.27  & (5.8 $\pm$ 0.1)$\cdot 10^{-2}$ \\
  36.79  & (5.1 $\pm$ 0.1)$\cdot 10^{-2}$ \\
  38.44  & (4.4 $\pm$ 0.1)$\cdot 10^{-2}$ \\
  40.25  & (3.91 $\pm$ 0.09)$\cdot 10^{-2}$ \\
  42.25  & (3.44 $\pm$ 0.08)$\cdot 10^{-2}$ \\
  44.47  & (3.03 $\pm$ 0.07)$\cdot 10^{-2}$ \\
  46.94  & (2.62 $\pm$ 0.06)$\cdot 10^{-2}$ \\
  49.71  & (2.23 $\pm$ 0.06)$\cdot 10^{-2}$ \\
  52.84  & (1.87 $\pm$ 0.05)$\cdot 10^{-2}$ \\
  56.40  & (1.56 $\pm$ 0.04)$\cdot 10^{-2}$ \\
  60.49  & (1.25 $\pm$ 0.03)$\cdot 10^{-2}$ \\
  65.23  & (1.04 $\pm$ 0.03)$\cdot 10^{-2}$ \\
  70.80  & (7.6 $\pm$ 0.2)$\cdot 10^{-3}$ \\
  77.42  & (6.3 $\pm$ 0.2)$\cdot 10^{-3}$ \\
  85.43  & (4.4 $\pm$ 0.1)$\cdot 10^{-3}$ \\
  95.34  & (3.2 $\pm$ 0.1)$\cdot 10^{-3}$ \\
 107.88  & (2.12 $\pm$ 0.08)$\cdot 10^{-3}$ \\
 124.27  & (1.42 $\pm$ 0.05)$\cdot 10^{-3}$ \\
 146.62  & (8.8 $\pm$ 0.4)$\cdot 10^{-4}$ \\
 178.85  & (4.8 $\pm$ 0.2)$\cdot 10^{-4}$ \\
 229.36  & (2.2 $\pm$ 0.1)$\cdot 10^{-4}$ \\
 319.72  & (7.5 $\pm$ 0.5)$\cdot 10^{-5}$ \\
 525.82  & (1.4 $\pm$ 0.1)$\cdot 10^{-5}$ \\
1288.74  & (5.9 $\pm$ 0.8)$\cdot 10^{-7}$ \\

\end{tabular}
\end{center}



\item M.P. De Pascale et al,  1993\cite{depascale1993}

          Positive Muons:     
\begin{center}
\begin{tabular}{cc}

  10.19  & (5.983 $\pm$ 0.233)$\cdot 10^{-1}$ \\ 
  14.42  & (2.523 $\pm$ 0.144)$\cdot 10^{-1}$ \\ 
  20.36  & (1.246 $\pm$ 0.071)$\cdot 10^{-1}$ \\
  28.80  & (4.709 $\pm$ 0.414)$\cdot 10^{-2}$ \\ 
  40.64  & (1.430 $\pm$ 0.162)$\cdot 10^{-2}$ \\ 
  70.16  & (5.176 $\pm$ 0.554)$\cdot 10^{-3}$ \\ 
\end{tabular}
\end{center}

           Negative Muons:
\begin{center}
\begin{tabular}{cc}
  10.19  & (5.025 $\pm$ 0.216)$\cdot 10^{-1}$ \\
  14.42  & (2.124 $\pm$ 0.132)$\cdot 10^{-1}$ \\
  20.36  & (8.653 $\pm$ 0.580)$\cdot 10^{-2}$ \\
  28.80  & (3.788 $\pm$ 0.375)$\cdot 10^{-2}$ \\
  40.64  & (1.389 $\pm$ 0.158)$\cdot 10^{-2}$ \\
  70.16  & (2.423 $\pm$ 0.383)$\cdot 10^{-3}$ \\

\end{tabular}
\end{center}

\item S. Tsuji et al,  1998\cite{tsuji1998}

\begin{center}
\begin{tabular}{cc}
 
12.1  & (8.37 $\pm$ 0.17)$\cdot 10^{-1}$ \\
 17.2  & (3.75 $\pm$ 0.12)$\cdot 10^{-1}$ \\
 22.3  & (2.04 $\pm$ 0.09)$\cdot 10^{-1}$ \\
 27.3  & (1.17 $\pm$ 0.07)$\cdot 10^{-1}$ \\
 34.3  & (6.12 $\pm$ 0.37)$\cdot 10^{-2}$ \\
 44.5  & (3.21 $\pm$ 0.28)$\cdot 10^{-2}$ \\
 54.6  & (1.39 $\pm$ 0.20)$\cdot 10^{-2}$ \\
 64.6  & (8.68 $\pm$ 1.64)$\cdot 10^{-3}$ \\
 74.7  & (8.07 $\pm$ 1.68)$\cdot 10^{-3}$ \\
 84.7  & (3.73 $\pm$ 1.18)$\cdot 10^{-3}$ \\
 94.7  & (1.69 $\pm$ 0.85)$\cdot 10^{-3}$ \\
119.8  & (1.00 $\pm$ 0.32)$\cdot 10^{-3}$ \\
171.2  & (2.84 $\pm$ 2.01)$\cdot 10^{-4}$ \\
222.0  & (3.59 $\pm$ 2.54)$\cdot 10^{-4}$ \\

\end{tabular}
\end{center}

\item J. Kremer et al,  1999\cite{kremer1999} 1994 data

          Positive Muons:  
\begin{center}
\begin{tabular}{cc}

  12.42  & (3.89 $\pm$ 0.08)$\cdot 10^{-1}$ \\ 
  18.85  & (1.38 $\pm$ 0.04)$\cdot 10^{-1}$ \\ 
  26.68  & (6.3 $\pm$ 0.3)  $\cdot 10^{-2}$ \\ 
  36.69  & (2.8 $\pm$ 0.1)  $\cdot 10^{-2}$ \\ 
  51.47  & (9.9 $\pm$ 0.7)  $\cdot 10^{-3}$ \\ 
  72.08  & (3.6 $\pm$ 0.3)  $\cdot 10^{-3}$ \\ 
 100.96  & (1.4 $\pm$ 0.2)  $\cdot 10^{-3}$ \\ 
\end{tabular}
\end{center}



\vspace*{0.4992cm}

          Negative Muons:
\begin{center}
\begin{tabular}{cc}
  12.42  & (3.09 $\pm$ 0.07)$\cdot 10^{-1}$ \\ 
  18.85  & (1.08 $\pm$ 0.03)$\cdot 10^{-1}$ \\
  26.68  & (4.6 $\pm$ 0.2)  $\cdot 10^{-2}$ \\
  36.69  & (1.9 $\pm$ 0.1)  $\cdot 10^{-2}$ \\
  51.47  & (7.1 $\pm$ 0.6)  $\cdot 10^{-3}$ \\
  72.08  & (3.0 $\pm$ 0.3)  $\cdot 10^{-3}$ \\
 100.96  & (1.2 $\pm$ 0.2)  $\cdot 10^{-3}$ \\

\end{tabular}
\end{center}

\item J. Kremer et al,  1999\cite{kremer1999} 1997 data

          Positive Muons:   
\begin{center}
\begin{tabular}{cc}

  12.42  & (4.14 $\pm$ 0.09)$\cdot 10^{-1}$ \\ 
  18.85  & (1.54 $\pm$ 0.04)$\cdot 10^{-1}$ \\ 
  26.68  & (6.4 $\pm$ 0.2)  $\cdot 10^{-2}$ \\ 
  36.69  & (2.8 $\pm$ 0.1)  $\cdot 10^{-2}$ \\ 
  51.47  & (10.2 $\pm$ 0.5) $\cdot 10^{-3}$ \\ 
  72.08  & (4.2 $\pm$ 0.3)  $\cdot 10^{-3}$ \\ 
 100.96  & (1.5 $\pm$ 0.1)  $\cdot 10^{-3}$ \\
\end{tabular}
\end{center}

          Negative Muons:
\begin{center}
\begin{tabular}{cc}

  12.42  & (3.20 $\pm$ 0.07)$\cdot 10^{-1}$ \\ 
  18.85  & (1.16 $\pm$ 0.03)$\cdot 10^{-1}$ \\
  26.68  & (4.5 $\pm$ 0.2)  $\cdot 10^{-2}$ \\
  36.69  & (2.03 $\pm$ 0.08)$\cdot 10^{-2}$ \\
  51.47  & (7.7 $\pm$ 0.4)  $\cdot 10^{-3}$ \\
  72.08  & (3.2 $\pm$ 0.2)  $\cdot 10^{-3}$ \\
 100.96  & (1.1 $\pm$ 0.1)  $\cdot 10^{-3}$ \\

\end{tabular}
\end{center}

\vspace*{8.5cm}

\vfill

\end{itemize}

\newpage

\underline{APPENDIX C}
Charge ratio data.

\medskip

The following lists contain all 
charge ratio data for momenta above $10 \, \mathrm{GeV}$. 
Each line contains the momentum 
in $\mathrm{GeV}$ together with the published
value and uncertainty of
the charge ratio.

\begin{itemize}
\item D.E. Caro et al, 1950\cite{caro1950}

\begin{center}
\begin{tabular}{cc}

35.0  & 1.6 $\pm$ 0.2
\end{tabular}
\end{center}

\item B.G. Owen et al, 1951\cite{owen1951}

\begin{center}
\begin{tabular}{cc}

  11.5  & 1.229 $\pm$ 0.036
\end{tabular}
\end{center}

\item I. Filosofo et al, 1954\cite{filosofo1954}

\begin{center}
\begin{tabular}{cc}

  21.0  & 1.232 $\pm$ 0.016
\end{tabular}
\end{center}

\item J. Pine et al, 1959\cite{pine1959}

\begin{center}
\begin{tabular}{cc}

  19.6       & 1.303 $\pm$ 0.031  \\ 
  22.8       & 1.29  $\pm$ 0.10   \\
  34.8       & 1.222 $\pm$ 0.052   \\
  48.6       & 1.15  $\pm$ 0.12   \\
\end{tabular}
\end{center}

\item J.E.R. Holmes et al, 1961\cite{holmes1961}

\begin{center}
\begin{tabular}{cc}

   6.7  & 1.39  $\pm$ 0.08 \\
  11.0  & 1.35  $\pm$ 0.08 \\
  18.0  & 1.29  $\pm$ 0.08 \\
  36.0  & 1.29  $\pm$ 0.14 \\
  98.0  & 1.02  $\pm$ 0.14 \\
\end{tabular}
\end{center}

\item W. Pak et al, 1961\cite{pak1961}

\begin{center}
\begin{tabular}{cc}

  13.1    & 1.252 $\pm$ 0.029 \\
  18.1    & 1.237 $\pm$ 0.064 \\
  25.3    & 1.262 $\pm$ 0.050 \\
  49.3    & 1.137 $\pm$ 0.093  \\
\end{tabular}
\end{center}


\vspace*{2cm}

\item P.J. Hayman and A.W. Wolfendale, 1962\cite{hayman1962}
 
\begin{center}
\begin{tabular}{cc}

  10.4  & 1.223 $\pm$ 0.038 \\
  17.5  & 1.233 $\pm$ 0.037 \\
  35.0  & 1.268 $\pm$ 0.051 \\
  77.0  & 1.37  $\pm$ 0.16 \\
 120.0  & 1.45  $\pm$ 0.23  \\
 240.0  & 1.51  $\pm$ 0.38  \\
\end{tabular}
\end{center}

%
%
%

\item I.C. Appleton et al, 1971\cite{appleton1971}

\begin{center}
\begin{tabular}{cc}

  12.7  & 1.312 $\pm$ 0.039 \\
  17.2  & 1.263 $\pm$ 0.038 \\
  28.3  & 1.306 $\pm$ 0.044 \\
  50.0  & 1.285 $\pm$ 0.085 \\
  81.0  & 1.165 $\pm$ 0.14 \\
 127.0  & 1.266 $\pm$ 0.20 \\
 288.0  & 1.105 $\pm$ 0.25 \\
\end{tabular}
\end{center}

\item O.C. Allkofer and W.D. Dau, 1972\cite{allkofer1972}
 
\begin{center}
\begin{tabular}{cc}

  11.4  & 1.22  $\pm$ 0.10
\end{tabular}
\end{center}

\item B.C. Nandi et al, 1972\cite{nandi1972}

\begin{center}
\begin{tabular}{cc}

  10.8  & 1.263 $\pm$ 0.030 \\
  15.2  & 1.268 $\pm$ 0.040 \\
  19.7  & 1.293 $\pm$ 0.057 \\
  26.6  & 1.290 $\pm$ 0.050 \\
  37.1  & 1.209 $\pm$ 0.079 \\
  46.9  & 1.257 $\pm$ 0.104 \\
  60.0  & 1.235 $\pm$ 0.114 \\
  84.0  & 1.430 $\pm$ 0.172 \\
 142.0  & 1.363 $\pm$ 0.188 \\
 260.0  & 1.364 $\pm$ 0.383 \\
 566.0  & 1.259 $\pm$ 0.460 \\
\end{tabular}
\end{center}

\vfill 

\break

\item J.M. Baxendale et al, 1975\cite{baxendale1975}

\begin{center}
\begin{tabular}{cc}

  11.5  & 1.259 $\pm$ 0.029 \\
  11.9  & 1.269 $\pm$ 0.029 \\
  12.3  & 1.229 $\pm$ 0.028 \\   
  12.8  & 1.371 $\pm$ 0.031 \\
  13.3  & 1.219 $\pm$ 0.027 \\
  13.9  & 1.261 $\pm$ 0.027 \\
  14.5  & 1.292 $\pm$ 0.028 \\
  15.3  & 1.291 $\pm$ 0.027 \\
  16.1  & 1.299 $\pm$ 0.027 \\
  17.0  & 1.256 $\pm$ 0.026 \\
  18.2  & 1.319 $\pm$ 0.028 \\
  19.5  & 1.279 $\pm$ 0.027 \\
  21.2  & 1.283 $\pm$ 0.021 \\
  22.7  & 1.300 $\pm$ 0.018 \\
  24.1  & 1.293 $\pm$ 0.035 \\
  25.2  & 1.301 $\pm$ 0.022  \\
  26.3  & 1.289 $\pm$ 0.036 \\
  27.8  & 1.285 $\pm$ 0.028 \\
  29.3  & 1.286 $\pm$ 0.036 \\
  31.5  & 1.325 $\pm$ 0.023  \\
  33.1  & 1.324 $\pm$ 0.037 \\
  35.3  & 1.300 $\pm$ 0.038 \\
  36.8  & 1.262 $\pm$ 0.028 \\  
  39.6  & 1.304 $\pm$ 0.019 \\
  43.2  & 1.283 $\pm$ 0.019 \\
  45.2  & 1.258 $\pm$ 0.024 \\
  47.1  & 1.222 $\pm$ 0.029 \\
  49.3  & 1.265 $\pm$ 0.031 \\
  50.7  & 1.358 $\pm$ 0.043 \\
  52.1  & 1.361 $\pm$ 0.034 \\
  54.9  & 1.250 $\pm$ 0.022  \\
  58.2  & 1.271 $\pm$ 0.025 \\
  63.0  & 1.223 $\pm$ 0.033 \\
  65.3  & 1.312 $\pm$ 0.046 \\
  68.1  & 1.272 $\pm$ 0.025  \\
  73.7  & 1.286 $\pm$ 0.037 \\
  76.6  & 1.323 $\pm$ 0.049 \\
  80.4  & 1.235 $\pm$ 0.028  \\
  88.3  & 1.215 $\pm$ 0.038 \\
  93.0  & 1.238 $\pm$ 0.050 \\
  98.3  & 1.245 $\pm$ 0.039 \\
 112.0  & 1.243 $\pm$ 0.044 \\
 118.0  & 1.270 $\pm$ 0.058 \\
 128.0  & 1.287 $\pm$ 0.047 \\
 145.0  & 1.268 $\pm$ 0.052 \\
\end{tabular}
\end{center}


\begin{center}
\begin{tabular}{cc}
 160.0  & 1.250 $\pm$ 0.068 \\
 177.0  & 1.221 $\pm$ 0.056 \\
 214.0  & 1.298 $\pm$ 0.071 \\
 236.0  & 1.327 $\pm$ 0.101 \\
 274.0  & 1.316 $\pm$ 0.093 \\
 358.0  & 1.586 $\pm$ 0.154 \\
\end{tabular}
\end{center}

\item B.C. Rastin, 1984\cite{rastin1984b}

\begin{center}
\begin{tabular}{cc}

  10.69  & 1.239 $\pm$ 0.016 \\
  11.94  & 1.247 $\pm$ 0.016 \\
  13.58  & 1.251 $\pm$ 0.016 \\
  15.81  & 1.285 $\pm$ 0.016 \\
  19.05  & 1.263 $\pm$ 0.016 \\
  24.14  & 1.233 $\pm$ 0.016 \\
  28.35  & 1.267 $\pm$ 0.053 \\
  29.30  & 1.166 $\pm$ 0.050 \\
  30.32  & 1.182 $\pm$ 0.051 \\
  31.42  & 1.250 $\pm$ 0.054 \\
  32.60  & 1.325 $\pm$ 0.058 \\
  33.88  & 1.253 $\pm$ 0.055 \\
  35.27  & 1.277 $\pm$ 0.057 \\
  36.79  & 1.238 $\pm$ 0.056 \\
  38.44  & 1.252 $\pm$ 0.059 \\
  40.25  & 1.206 $\pm$ 0.057 \\
  42.25  & 1.277 $\pm$ 0.060 \\
  44.47  & 1.291 $\pm$ 0.062 \\
  46.94  & 1.388 $\pm$ 0.069 \\
  49.71  & 1.289 $\pm$ 0.064 \\
  52.84  & 1.160 $\pm$ 0.059 \\
  56.40  & 1.273 $\pm$ 0.067 \\
  60.49  & 1.240 $\pm$ 0.067 \\
  65.23  & 1.310 $\pm$ 0.073 \\
  70.80  & 1.157 $\pm$ 0.066 \\
  77.42  & 1.203 $\pm$ 0.071 \\
  85.43  & 1.256 $\pm$ 0.080 \\
  95.34  & 1.207 $\pm$ 0.082 \\
 107.88  & 1.320 $\pm$ 0.096 \\
 124.27  & 1.194 $\pm$ 0.091 \\
 146.62  & 1.235 $\pm$ 0.101 \\
 178.85  & 1.161 $\pm$ 0.104 \\
 229.36  & 1.234 $\pm$ 0.127 \\
 319.72  & 1.30  $\pm$ 0.16 \\
 525.82  & 1.32  $\pm$ 0.22 \\
1288.74  & 1.14  $\pm$ 0.29 \\
\end{tabular}
\end{center}

\vspace*{1cm}

\item M.P. De Pascale et al,  1993\cite{depascale1993}

\begin{center}
\begin{tabular}{cc}

  19.89  & 1.292 $\pm$ 0.075 \\
  55.87  & 1.409 $\pm$ 0.173 \\
\end{tabular}
\end{center}




\item J. Kremer et al,  1999\cite{kremer1999} 1994 data

\begin{center}
\begin{tabular}{cc}
  12.42  & 1.257 $\pm$ 0.038 \\
  18.85  & 1.269 $\pm$ 0.055 \\
  26.68  & 1.372 $\pm$ 0.086 \\
  36.69  & 1.466 $\pm$ 0.110 \\
  51.47  & 1.384 $\pm$ 0.144 \\
  72.08  & 1.212 $\pm$ 0.169 \\
 100.96  & 1.235 $\pm$ 0.233 \\
\end{tabular}
\end{center}

\item J. Kremer et al,  1999\cite{kremer1999} 1997 data

\begin{center}
\begin{tabular}{cc}

  12.42  & 1.291 $\pm$ 0.027 \\
  18.85  & 1.335 $\pm$ 0.038 \\
  26.68  & 1.427 $\pm$ 0.058 \\
  36.69  & 1.383 $\pm$ 0.068 \\
  51.47  & 1.320 $\pm$ 0.090 \\
  72.08  & 1.337 $\pm$ 0.120 \\
 100.96  & 1.337 $\pm$ 0.169 \\
\end{tabular}
\end{center}

\vfill

\end{itemize}

\end{multicols}


\vspace*{2cm}

\underline{APPENDIX D} 
Correlation matrix for the fit of the spectrum shape:

%
\begin{center}
\begin{tabular}{|l|llll|}
\hline
       & $H_3$  &  $H_2$ & $H_1$  & $S_2$ \\ \hline
 $H_3$ &  1.000 &  0.319 & -0.298 & -0.771 \\
 $H_2$ &  0.319 &  1.000 & -0.105 & -0.558 \\
 $H_1$ & -0.298 & -0.105 & 1.000 &  0.672 \\
 $S_2$ & -0.771 & -0.558 &  0.672 & 1.000 \\ \hline

\end{tabular}
\end{center}

\end{document}